%% file: HIG-14-036_temp.tex
\pdfoutput=1

\documentclass[11pt,twoside,a4paper,cmspaper,final,collab]{cms-tdr}
\def\svnVersion{310178}\def\cmsCernNo{2015-159}\def\cmsCernDate{\today}\def\cmsMessage{Submitted to Physical Review D}

\begin{document}\cmsNoteHeader{HIG-14-036}

\hyphenation{had-ron-i-za-tion}
\hyphenation{cal-or-i-me-ter}
\hyphenation{de-vices}
\RCS$Revision: 305585 $
\RCS$HeadURL: svn+ssh://alverson@svn.cern.ch/reps/tdr2/papers/HIG-14-036/trunk/HIG-14-036.tex $
\RCS$Id: HIG-14-036.tex 305585 2015-10-09 23:51:26Z usarica $
\newlength\cmsFigWidth
\ifthenelse{\boolean{cms@external}}{\setlength\cmsFigWidth{0.48\textwidth}}{\setlength\cmsFigWidth{0.7\textwidth}}
\ifthenelse{\boolean{cms@external}}{\providecommand{\cmsLeft}{top\xspace}}{\providecommand{\cmsLeft}{left\xspace}}
\ifthenelse{\boolean{cms@external}}{\providecommand{\cmsRight}{bottom\xspace}}{\providecommand{\cmsRight}{right\xspace}}
\newcommand{\GH}{\ensuremath{\Gamma_{\PH}}\xspace}
\newcommand{\GHSM}{\ensuremath{\Gamma_{\PH}^{\mathrm{SM}}}\xspace}
\newcommand{\V}{\ensuremath{\cmsSymbolFace{V}}\xspace}
\newcommand{\fLQ}{\ensuremath{f_{\Lambda Q}}\xspace}
\newcommand{\VDlife}{\ensuremath{\Delta t}\xspace}
\newcommand{\phiLQ}{\ensuremath{\phi_{\Lambda Q}}\xspace}
\newcommand{\mell}{\ensuremath{m_{4\ell}}\xspace}
\newcommand{\ttH}{\ensuremath{\ttbar\PH}\xspace}
\providecommand{\mH}{\ensuremath{m_{\PH}}\xspace}
\ifthenelse{\boolean{cms@external}}{\providecommand{\CLns}{C.L\xspace}}{\providecommand{\CLns}{CL\xspace}}
\ifthenelse{\boolean{cms@external}}{\providecommand{\CL}{C.L.\xspace}}{\providecommand{\CL}{CL\xspace}}
\providecommand{\CLs}{\ensuremath{\mathrm{CL}_\mathrm{s}}\xspace}

\cmsNoteHeader{HIG-14-036}
\title{Limits on the Higgs boson lifetime and width from its decay to four charged leptons}

\date{\today}

\abstract{
Constraints on the lifetime and width of the Higgs boson are obtained from
$\PH\to \Z\Z \to 4\ell$ events
using data recorded by the CMS experiment during the LHC run 1 with an integrated
luminosity of 5.1 and 19.7\fbinv at a center-of-mass energy of 7 and 8\TeV, respectively.
The measurement of the Higgs boson lifetime is derived from its flight distance
in the CMS detector with an upper bound of $\tau_{\PH}<1.9\times10^{-13}$\unit{s} at the 95\% confidence level (\CL),
corresponding to a lower bound on the width of $\GH > 3.5\times10^{-9}$\MeV.
The measurement of the width is obtained from an off-shell production technique,
generalized to include anomalous couplings of the Higgs boson to two electroweak bosons.
From this measurement, a joint constraint is set on the Higgs boson width and a parameter $\fLQ$ that
expresses an anomalous coupling contribution as an on-shell cross-section fraction.
The limit on the Higgs boson width is $\GH<46$\MeV with $\fLQ$ unconstrained
and $\GH<26$\MeV for $\fLQ=0$ at the 95\% \CLns.
The constraint $\fLQ < 3.8\times10^{-3}$ at the 95\% \CL is obtained for the expected standard model Higgs boson width.
}

\hypersetup{%
pdfauthor={CMS Collaboration},%
pdftitle={Limits on the Higgs boson lifetime and width from its decay to four charged leptons},%
pdfsubject={CMS},%
pdfkeywords={CMS, physics, Higgs, four-lepton, decay, four charged leptons, lifetime, width, anomalous coupling}}

\maketitle

\section{Introduction} \label{sec:Introduction}

The discovery of a new boson with mass of about 125\GeV
by the ATLAS and CMS experiments~\cite{Aad:2012tfa, Chatrchyan:2012ufa, Chatrchyan:2013lba}
at the CERN LHC provides support for the standard model (SM) mechanism with a field responsible for generating the
masses of elementary particles~\cite{Englert:1964et,Higgs:1964pj,Guralnik:1964eu,
StandardModel67_1,StandardModel67_2,StandardModel67_3}.
This new particle is believed to be a Higgs boson (\PH), the scalar particle appearing as an excitation of this field.
The measurement of its properties, such as the lifetime, width, and structure of its couplings to the known SM particles, is of high priority to determine its nature.

The CMS and ATLAS experiments have set constraints of
$\GH < 22\MeV$ at 95\% confidence level (\CL) on the \PH boson total width~\cite{CMS-HIG-14-002,Aad:2015xua} from the ratio of off-shell to on-shell production.
The precision on $\GH$ from direct on-shell measurements alone is approximately 1\GeV~\cite{Khachatryan:2014jba,Aad:2014aba}, which is significantly larger.
The two experiments have also set constraints on the spin-parity properties and anomalous couplings of the \PH
boson~\cite{Chatrchyan:2012jja,Aad:2013xqa,Chatrchyan:2013mxa,Khachatryan:2014kca,Aad:2015rwa},
finding its quantum numbers to be consistent with $J^{PC}=0^{++}$ but allowing small anomalous
coupling contributions. No direct experimental limit on the \PH boson lifetime was set,
and the possible presence of anomalous couplings was not considered in the constraints on the \PH boson width.
This paper provides these two measurements.

The measurement of the \PH boson lifetime in this paper is derived from its flight distance
in the CMS detector~\cite{CMSDETECTOR}, and
the measurement of the width is obtained from the off-shell production technique,
generalized to include anomalous couplings of the \PH boson to two electroweak bosons, $\PW\PW$ and $\Z\Z$.
From the latter measurement, a joint constraint is set on the \PH boson width and a parameter that
quantifies an anomalous coupling contribution as on-shell cross-section fraction.
The event reconstruction and analysis techniques rely on the previously published results \cite{Chatrchyan:2013mxa,CMS-HIG-14-002,Khachatryan:2014kca,Khachatryan:2015cwa},
and their implementations are discussed in detail. Only the final state with four charged leptons is considered in this paper, but the constraints on the width could be improved by including final states with
neutrinos in the off-shell production~\cite{CMS-HIG-14-002,Aad:2015xua}.
Indirect constraints on the \PH boson width and lifetime are also possible through the combination
of data on \PH boson production and decay rates~\cite{Khachatryan:2014jba, Aad:2013wqa}.
While such a combination tests the compatibility of the data with the SM \PH boson, it relies on stronger theoretical assumptions such as SM-like coupling ratios among the different final states.

Section~\ref{sec:pheno} in this paper discusses the analysis methods for measuring the \PH boson lifetime, and for relating the anomalous couplings of the \PH boson to the measurement of $\GH$ through the off-shell production technique. Section~\ref{sec:data} discusses the CMS detector and event simulation, and Sec.~\ref{sec:selection} defines the selection criteria used in the analysis. Section~\ref{sec:observables} describes the analysis observables, categorization, and any related uncertainty. Section~\ref{sec:lifetime} provides the constraints on the \PH boson lifetime, while Sec.~\ref{sec:width} provides the upper limits for both the \PH boson width and the anomalous coupling parameter investigated in this paper. The summary of results is provided in Sec.~\ref{sec:Conclusions}.

\section{Analysis techniques} \label{sec:pheno}

The lifetime of each $\PH$ boson candidate in its rest frame is determined in a four-lepton event as
\begin{equation}
\VDlife = \frac{\mell}{\pt}    \left(\Delta\vec{r}_\mathrm{T}  \cdot \hat{p}_\mathrm{T}\right),
\label{eq:tau_define}
\end{equation}
where  $\mell$ is the four-lepton invariant mass, $\Delta\vec{r}_\mathrm{T}$ is the displacement vector between the decay vertex and the production vertex of the \PH boson in the plane transverse
to the beam axis,
and $\hat{p}_\mathrm{T}$ and \pt are respectively the unit vector and the magnitude of the \PH boson transverse momentum.
The average $\VDlife$ is inversely proportional to the total width:
\begin{equation}
\langle \VDlife \rangle = \tau_\PH = \frac{\hbar}{\GH}
\label{eq:tauHgammaH}
\end{equation}
The distribution of the measured lifetime $\VDlife$ is used to set an upper limit on the average lifetime of the
\PH boson, or equivalently a lower limit on its width $\GH$, and it follows the exponential distribution if known perfectly. The expected SM \PH boson average lifetime is $\tau_\PH \approx 48$\unit{fm/c} ($16\times10^{-8}$\unit{fs})
and is beyond instrumental precision.
The technique summarized in Eq.~(\ref{eq:tau_define}) nonetheless
allows the first direct experimental constraint on~$\tau_\PH$.

The upper bound on $\GH$ is set using the off-shell production method~\cite{CaolaMelnikov:1307.4935,Kauer:2012hd,Campbell:2013una}
and follows the technique developed by CMS~\cite{CMS-HIG-14-002}, where the gluon fusion and weak vector boson fusion
(VBF) production mechanisms were considered in the analysis. The technique considers the \PH boson production relationship between the on-shell
($105.6 < \mell< 140.6$\GeV) and off-shell ($220 < \mell< 1600$\GeV) regions.
Denoting each production mechanism with
$\text{vv} \to \PH \to \Z\Z$ for \PH boson coupling to either strong ($\text{vv}=\Pg\Pg$)
or weak ($\text{vv}=\V\V$) vector bosons $\text{vv}$,
the on-shell and off-shell yields are related by
\begin{equation}
\label{eq:resonant}
\sigma^\text{on-shell}_{\text{vv} \to \PH \to \Z\Z} \propto \mu_{\text{vv}\PH}
\text{  and  }
\sigma^\text{off-shell}_{\text{vv}  \to \PH \to \Z\Z} \propto \mu_{\text{vv}\PH}\, \GH,
\end{equation}
where $\mu_{\text{vv}\PH}$ is the on-shell signal strength, the ratio of the observed and expected on-shell production cross sections for the four-lepton final state, which is denoted by either $\mu_{\Pg\Pg\PH}$ for gluon fusion production
or
$\mu_{\V\V\PH}$ for VBF production.
The $\ttbar\PH$ process is driven by the \PH boson couplings to heavy quarks like the gluon fusion process,
and
the $\V\PH$ process by the \PH boson couplings to weak vector bosons like the VBF process.
They are therefore parametrized with the same on-shell signal strengths
$\mu_{\Pg\Pg\PH}$ and $\mu_{\V\V\PH}$, respectively.
The effects of signal-background
interference are not shown in Eq.~(\ref{eq:resonant}) for illustration but are taken into account in the analysis.

The relationship in Eq.~(\ref{eq:resonant}) implies variations of the $\text{vv}\PH$ couplings as a function of $\mell$.
This variation is assumed to be as in the SM in the gluon fusion process. The assumption is valid as long as the production is dominated
by the top-quark loop and no new particles contribute to this loop.
Variation of the $\PH\V\V$ couplings, either in the VBF or $\V\PH$ production or in the $\PH\to\Z\Z$ decay,
may depend on anomalous coupling contributions. An enhancement of the off-shell signal production is
suggested with anomalous $\PH\V\V$ couplings~\cite{CMS-HIG-14-002, Gainer:2014hha, Englert:2014aca, Ghezzi:2014qpa}, but
neither experimental studies of off-shell production nor realistic treatment of signal-background interference has been done with these anomalous couplings.
We extend the methodology of the recent analysis of
anomalous $\PH\V\V$ couplings of the \PH boson~\cite{Khachatryan:2014kca}
to study these couplings and introduce in the scattering amplitude an additional term that depends on
the \PH boson invariant mass, $\left(q_{{\V}1} + q_{{\V}2}\right)^{2}$:
\ifthenelse{\boolean{cms@external}}{
\begin{multline}
\label{eq:fullampl-formfact-spin0}
A(\PH\V\V) \propto
\Biggl[ a_{1}
 - \re^{i\phi_{\Lambda{Q}}} \frac{\left(q_{{\V}1} + q_{{\V}2}\right)^{2}}{\left(\Lambda_{Q}\right)^{2}}\\
 - \re^{i\phi_{\Lambda{1}}} \frac{\left(q_{{\V}1}^2 + q_{{\V}2}^2\right)}{\left(\Lambda_{1}\right)^{2}}
\Biggr]m_{\V}^2 \epsilon_{{\V}1}^* \epsilon_{{\V}2}^*
\\
+ a_{2}^{}  f_{\mu \nu}^{*(1)}f^{*(2),\mu\nu}
+ a_{3}^{}   f^{*(1)}_{\mu \nu} {\tilde f}^{*(2),\mu\nu},
\end{multline}
}{
\begin{multline}
\label{eq:fullampl-formfact-spin0}
A(\PH\V\V) \propto
\left[ a_{1}
 - \re^{i\phi_{\Lambda{Q}}} \frac{\left(q_{{\V}1} + q_{{\V}2}\right)^{2}}{\left(\Lambda_{Q}\right)^{2}}
 - \re^{i\phi_{\Lambda{1}}} \frac{\left(q_{{\V}1}^2 + q_{{\V}2}^2\right)}{\left(\Lambda_{1}\right)^{2}}
 \right]
m_{\V}^2 \epsilon_{{\V}1}^* \epsilon_{{\V}2}^*
\\
+ a_{2}^{}  f_{\mu \nu}^{*(1)}f^{*(2),\mu\nu}
+ a_{3}^{}   f^{*(1)}_{\mu \nu} {\tilde f}^{*(2),\mu\nu},
\end{multline}
}
where $f^{(i){\mu \nu}} = \epsilon_{{\V}i}^{\mu}q_{{\V}i}^{\nu} - \epsilon_{{\V}i}^\nu q_{{\V}i}^{\mu} $
is the field strength tensor of a gauge boson with momentum $q_{{\V}i}$ and polarization
vector $\epsilon_{{\V}i}$, ${\tilde f}^{(i)}_{\mu \nu} = \frac{1}{2} \epsilon_{\mu\nu\rho\sigma} f^{(i),\rho\sigma}$
is the dual field strength tensor, the superscript~$^*$ designates a complex conjugate, and
$m_{\V}$ is the pole mass of a vector boson.
The $a_i$ are complex coefficients, and
the $\Lambda_{1}$ or $\Lambda_{Q}$ may be interpreted as the scales of beyond-the-SM (BSM) physics.
The complex phase of the $\Lambda_{1}$ and $\Lambda_{Q}$ terms are explicitly given as $\phi_{\Lambda{1}}$ and $\phi_{\Lambda{Q}}$, respectively.
Equation~(\ref{eq:fullampl-formfact-spin0}) describes all anomalous contributions up to dimension five operators.
In the SM, only the $a_{1}$ term appears at tree level in couplings to $\Z\Z$ and $\PW\PW$, and it remains dominant after loop corrections.
Constraints on the anomalous contributions from the $a_2$, $a_3$ and $\Lambda_{1}$ terms to the $\PH\to\V\V$ decay
have been set by the CMS and ATLAS experiments~\cite{Chatrchyan:2013mxa,Khachatryan:2014kca,Aad:2015rwa}
through on-shell \PH boson production.

The $\Lambda_{Q}$ term depends only on the invariant mass of the \PH boson, so its contribution is not distinguishable from the SM in the on-shell region. This paper tests the $\Lambda_{Q}$ term through the off-shell region.
Equation~(\ref{eq:fullampl-formfact-spin0}) describes both $\Z\Z$ and $\PW\PW$ couplings, and
it is assumed that $\Lambda_{Q}$ is the same for both.
The ratio of any loop contribution from a heavy particle in the $\PH\V\V$ scattering amplitude to the SM tree-level $a_1$ term would be predominantly real, and the imaginary part of the ratio would be small. If the contribution instead comes from an additional term to the SM Lagrangian itself, this ratio can only be real.
Therefore, only real coupling ratios are tested such that
$\cos\phi_{\Lambda{Q}}=\pm1$ and $a_{1} \ge 0$, where $a_{1}=2$ and $\Lambda_{Q} \to \infty$ correspond to the tree-level SM $\PH\V\V$ scattering with $\mu_{\Pg\Pg\PH}=\mu_{\V\V\PH}=1$.
The effective cross-section fraction due to the $\Lambda_{Q}$ term, denoted as \fLQ, allows a parametrization similar to the conventions of $\Lambda_{1}$ in Ref.~\cite{Khachatryan:2014kca}. It is defined for the on-shell $\Pg\Pg\to\PH\to\V\V$ process assuming no contribution from other anomalous couplings as
\begin{equation}
\fLQ = \frac{m^4_\PH/ \Lambda_Q^4}{\abs{a_{1}}^2 + m^4_\PH/ \Lambda_Q^4}.
\label{eq:fLQ_definition}
\end{equation}

The $\PH\V\V$ couplings in Eq.~(\ref{eq:fullampl-formfact-spin0}) appear in both production
and decay for the VBF and $\V\PH$ mechanisms while they appear only in decay for \PH boson production through gluon fusion. Isolating the former two production mechanisms, therefore, enhances the sensitivity to the contribution of
anomalous couplings.
While the previous study of the \PH boson width~\cite{CMS-HIG-14-002} employs dijet tagging
only in the on-shell region, VBF jet identification is also extended to
the off-shell region in this analysis with techniques from Ref.~\cite{Khachatryan:2015cwa}. A joint constraint is obtained
on $\GH$, $\fLQ$, $\mu_{\Pg\Pg\PH}$, and $\mu_{\V\V\PH}$, where the latter two parameters correspond to
the $\PH$ production strength in gluon fusion, and VBF or $\V\PH$ production mechanisms in the on-shell region, respectively.

\section{The CMS experiment and simulation} \label{sec:data}

The CMS detector, described in detail in Ref.~\cite{CMSDETECTOR}, provides excellent resolution
for the measurement of electron and muon momenta and impact parameters near the
LHC beam interaction region.
Within the superconducting solenoid (3.8\unit{T}) volume of CMS, there are a silicon pixel and strip tracker,
a lead tungstate crystal electromagnetic calorimeter (ECAL) and a brass and scintillator hadron calorimeter.
Muons are identified in gas-ionization detectors embedded in the iron flux return placed outside the solenoid.
The data samples used in this analysis are the same as those described in
Refs.~\cite{Chatrchyan:2013mxa,CMS-HIG-14-002,Khachatryan:2014kca,Khachatryan:2015cwa},
corresponding to an integrated luminosity of 5.1\fbinv collected in proton-proton collisions at LHC
with center-of-mass energy of 7\TeV in 2011 and 19.7\fbinv at 8\TeV in 2012.
The uncertainties in the integrated luminosity measurement are 2.2\% and 2.6\% for the 2011 and 2012 data sets, respectively~\cite{CMS:2012eui,CMS-PAS-LUM-13-001}.

The \PH boson signal production through gluon fusion or in association with two fermions
from either vector boson fusion or associated vector boson production may interfere with the background $4\ell$ production with the same initial and final states. The background $4\ell$ production is considered to be any process that does not include a contribution from the \PH boson signal.
The on-shell Monte Carlo (MC) simulation does not require interference with the background
because of the relatively small \PH boson width~\cite{CMS-HIG-14-002}.
The off-shell production leads to a broad $\mell$ spectrum and
is generated using the full treatment of the interference between the signal and
background for each production mechanism.
Therefore, different techniques and tools have been used for on-shell and off-shell simulation.
The simulation of the \PH boson signal is
performed at the measured value of the \PH boson pole mass $\mH = 125.6\GeV$ in the $4\ell$ final state~\cite{Chatrchyan:2013mxa},
and the expected SM \PH boson width
$\GHSM = 4.15\MeV$~\cite{LHCHiggsCrossSectionWorkingGroup:2011ti,Heinemeyer:2013tqa}
along with several other \GH reference values.

The two dominant \PH boson production mechanisms, gluon fusion and VBF,
are generated on-shell at next-to-leading order (NLO) in perturbative quantum chromodynamics (QCD) using the
\POWHEG~\cite{Frixione:2007vw,Bagnaschi:2011tu,Nason:2009ai} event generator. The decay of the
\PH boson via $\PH\to\Z\Z\to4\ell$, including interference effects of identical leptons in the final state and nonzero
lifetime of the \PH boson, is modeled with \textsc{JHUGen}~4.8.1~\cite{Gao:2010qx,Bolognesi:2012mm,Anderson:2013afp}.
In addition, gluon fusion production with up to two jets at NLO in QCD has been generated
using \POWHEG with the \textsc{hjj} program~\cite{Campbell:2012am},
where the \textsc{minlo} procedure~\cite{Hamilton:2012np} is used to resum all large logarithms associated with
the presence of a scale for merging the matrix element and the parton shower contributions.
In all of the above cases, simulations with a wide range of masses $m_\PH$ up to 1000\GeV~\cite{Khachatryan:2015cwa} for \PH boson on-shell signal production at NLO in QCD
have been used to calibrate the behavior of associated particles in the simulation of off-shell \PH boson signal at leading order (LO) in QCD, which is described below.
The $\V\PH$ and $\ttbar\PH$ production mechanisms of the \PH boson, which have the smallest expected cross sections,
and the subsequent \PH boson prompt decays are simulated on-shell using \PYTHIA~6.4.24~\cite{Sjostrand:2006za}.

Four different values of the \PH boson lifetime have been generated with $c\tau_\PH = 0$, 100, 500, 1000\mum
for the gluon fusion production mechanism, and these samples are reweighted to model the values of lifetime in between the generated values. The only difference between gluon fusion and the other production mechanisms relevant for the constraint on the lifetime is the \PH boson $\pt$ spectrum, so
reweighting as a function of $\pt$ allows the modeling of the different production mechanisms with nonzero \PH boson lifetime.
Following the formalism in Eq.~(\ref{eq:fullampl-formfact-spin0}) for spin-zero
and including nonzero spin hypotheses, \textsc{JHUGen} simulations for a variety of \PH boson production (gluon fusion, VBF, $\V\PH$, $\ttH$, $\qqbar$) and decay
($\PH\to \Z\Z/\Z\gamma^*/\gamma^*\gamma^*\to 4\ell$) modes have been generated
with SM and BSM couplings to validate model independence of the lifetime analysis. This simulation is detailed in Ref.~\cite{Khachatryan:2014kca}.

The off-shell \PH boson signal and the interference effects with the background are included at LO in QCD for
gluon fusion, VBF, and $\V\PH$ mechanisms, while the $\ttbar\PH$ production is highly suppressed at higher
masses and is therefore not simulated off-shell~\cite{LHCHiggsCrossSectionWorkingGroup:2011ti,Heinemeyer:2013tqa}.
On-shell and off-shell events from gluon fusion production are generated with the  \MCFM~6.7~\cite{MCFM,Campbell:2011bn,Campbell:2013una} and
\textsc{gg2VV}~3.1.5~\cite{Binoth:2008pr} MC generators while those for the VBF and
associated production with an electroweak boson $\V$ are generated with \textsc{phantom}~1.2.3~\cite{Ballestrero:2007}.
The leptonic decay of the associated $\V$ boson is modeled with a reweighting procedure based on the branching ratios
of the $\V$ boson~\cite{Agashe:2014kda}, and the relatively small contribution of  $\PH\PH$ production is removed
from the \textsc{phantom} simulation.
Pure signal, pure background, and several mixed samples with signal-background interference
have been produced for the analysis of the interference effects.
The modeling of the anomalous couplings from Eq.~(\ref{eq:fullampl-formfact-spin0})
in the off-shell \PH boson production is performed by reweighting the SM-like samples.
An extended \MCFM library provided as part of the Matrix Element Likelihood Approach (MELA)~package, \cite{Gao:2010qx,Bolognesi:2012mm,Anderson:2013afp},
allows for both reweighting and event simulation with anomalous couplings in off-shell \PH production, and the analytical reweighting for the $\fLQ$ parametrization used in this analysis is identical to reweighting via the MELA package.

\begin{figure}[tbh]
\centering
\includegraphics[width=0.48\textwidth]{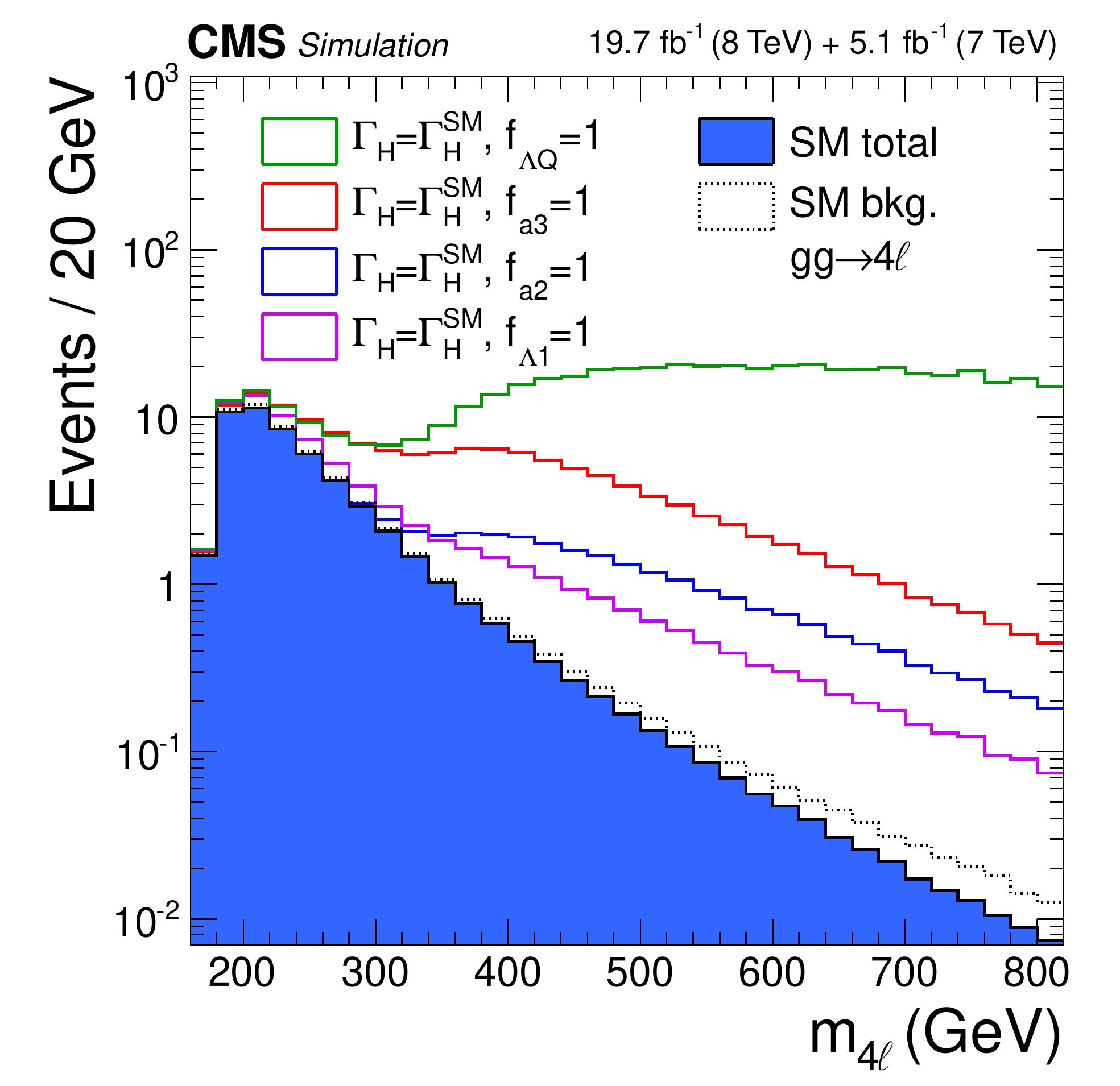}
\caption{
The $\mell$ distributions in the off-shell region in the simulation of the $\Pg\Pg \to 4\ell$ process with the $\Lambda_{Q}$ ($\fLQ=1$), $a_3$ ($f_{a3}=1$), $a_2$ ($f_{a2}=1$), and $\Lambda_{1}$ ($f_{\Lambda 1}=1$) terms, as open histograms, as well as the $a_1$ term (SM), as the filled histogram, from Eq.~(\ref{eq:fullampl-formfact-spin0}) in decreasing order of enhancement at high $\mell$. The on-shell signal yield and the width $\GH$ are constrained to the SM expectations. In all cases, the background and its interference with different signal hypotheses are included except in the case of the pure background (dotted), which has greater off-shell yield than the SM signal-background contribution due to destructive interference.
}
\label{fig:MCFM_BSM_GenLevel}

\end{figure}

Figure~\ref{fig:MCFM_BSM_GenLevel} illustrates the simulation of the $\Pg\Pg \to 4\ell$ process with the above
technique, which includes $\PH$ boson off-shell production, its background, and their interference for the five signal models with the $a_1$ (SM), $a_2$, $a_3$, $\Lambda_{1}$, and $\Lambda_{Q}$ terms in Eq.~(\ref{eq:fullampl-formfact-spin0}).
In all cases, the on-shell yield and the width $\GH$ are constrained to the SM expectations, and large enhancements are seen in the off-shell region.
The four BSM models correspond to the effective fractions $f_{ai}=1$ defined in
Ref.~\cite{Khachatryan:2014kca} or Eq.~(\ref{eq:fLQ_definition}).
When the on-shell contributions of the anomalous couplings are small, cancellation effects in the off-shell region due to their interference with the $a_1$ term or with the background, as in the case of the $\Lambda_{Q}$ term, may suppress the off-shell yield for a given $\GH$.
Among these four BSM models, the $\Lambda_{Q}$ term results in the largest off-shell enhancement,
and only the $\Lambda_{Q}$ and $a_1$ terms, and their interference between each other and
the background are considered in the width analysis.
Constraints on the $a_2$, $a_3$, and $\Lambda_{1}$ terms have already been measured from on-shell analyses~\cite{Khachatryan:2014kca,Aad:2015rwa}.

In the case of the off-shell MC simulation, the QCD renormalization and factorization scales are set to
the dynamic scales $\mell/2$ for gluon fusion and $\mell$ for
the VBF+$\V\PH$ signal productions and their backgrounds.
Higher-order QCD corrections for the gluon fusion signal process
are known to an accuracy of next-to-next-to-leading order (NNLO) and next-to-next-to-leading logarithms
for the total cross section~\cite{LHCHiggsCrossSectionWorkingGroup:2011ti,Heinemeyer:2013tqa},
and to NNLO as a function of $\mell$~\cite{Passarino:1312.2397v1}.
The $\mell$-dependent correction factors to the LO cross section (K factors) are typically in the range of 2.0 to 2.7.
Although no exact calculation exists beyond the LO for the $\cPg\cPg \to \Z\Z$ continuum background,
it has been recently shown~\cite{Bonvini:1304.3053}
that the soft collinear approximation is able to describe the background cross section and the
interference term at NNLO\@. Further calculations also show that the K factors are very similar at NLO for the signal and background~\cite{Melnikov:2015laa} and at NNLO for the signal and interference terms~\cite{Li:2015jva}. Therefore, the same K factor is used for the signal and background~\cite{Passarino:1312.2397v1}.
Similarly, QCD and electroweak corrections are known to an accuracy of NNLO for the VBF and $\V\PH$ signal
contributions~\cite{LHCHiggsCrossSectionWorkingGroup:2011ti,Heinemeyer:2013tqa,Brein:1111.0761},
but no calculation exists beyond the LO for the corresponding background contributions.
The same K factors as in signal are also assumed for the background and interference contributions.
Uncertainties due to the limited theoretical knowledge of the background K factor have a small impact on the final results.

The background $\Pq \Paq \to \Z\Z$ process is simulated using \POWHEG at NLO in QCD with no interference with \PH boson signal production. The NLO electroweak calculations~\cite{Bierweiler:1312,Baglio:1307.4331}
predict negative, $\mell$-dependent corrections to this
process for on-shell $\Z$ boson pairs and are taken into account. In addition, a two-jet inclusive
\MADGRAPH 5.1.3.30 ~\cite{Alwall:2007st} simulation is used to check jet categorization in the $\Pq \Paq \to \Z\Z$ process.
\PYTHIA is used to simulate parton showering and hadronization for all MC signal and background events. The generated MC events are subsequently processed with the CMS
full detector simulation, based on \GEANTfour~\cite{Agostinelli2003250}, and reconstructed using the same
algorithm used for the events in data.

The background from $\Z$ production with associated jets, denoted as $\Z+X$,
comes from the production of $\cPZ$ and $\PW\cPZ$ bosons in association with jets
as well as from $\ttbar$ production with one or two jets misidentified as an electron or a muon.
The estimation of the $\Z+X$ background in the four-lepton final state is obtained from data control regions without relying on simulation~\cite{Chatrchyan:2013mxa}.

\section{Event selection} \label{sec:selection}

The event reconstruction and selection requirements are the same as those in the previous measurements
of the \PH boson properties in the $\PH\to4\ell$
channel~\cite{Chatrchyan:2013mxa,CMS-HIG-14-002,Khachatryan:2014kca,Khachatryan:2015cwa}.
Only small modifications are made to the lepton impact parameter requirements in the lifetime
analysis to retain potential signal with a displaced four-lepton vertex.

As in previous measurements \cite{Chatrchyan:2013mxa,CMS-HIG-14-002,Khachatryan:2014kca,Khachatryan:2015cwa}, events are triggered by requiring the presence of two leptons (electrons or muons) with asymmetric
requirements on their $\pt$. A triple-electron trigger is also used.
Electron candidates are defined by a reconstructed charged-particle track in the
tracker pointing to an energy deposition in the ECAL.
A muon candidate is identified as a charged-particle track in the muon system
that matches a track reconstructed in the tracker.
The electron energy is measured primarily from the ECAL cluster energy, while the muon momentum
and the charged-lepton impact parameters near the interaction region
are measured primarily by the tracker.
Electrons and muons are required to be isolated from other charged and neutral particles~\cite{Chatrchyan:2013mxa}.
Electrons (muons) are reconstructed for $\pt > 7\,(5)\GeV$ within the geometrical
acceptance $\abs{\eta} < 2.5\,(2.4)$~\cite{Khachatryan:2015hwa,Chatrchyan:2012xi}.
Trigger and reconstruction efficiencies for muons and electrons are found to be
independent on the lifetime of the \PH boson, similar to other studies of long-lived
particles~\cite{CMS:2014hka,Khachatryan:2014mea}.

Events are selected with at least four identified and isolated electrons or muons to form the four-lepton candidate.
Two $\Z\to \ell^+\ell^-$ candidates originating from a pair of leptons of the same flavor and
opposite charge are required.
The $\ell^+\ell^-$ pair with an invariant mass, $m_1$, nearest to the nominal $\Z$ boson mass
is denoted $\Z_{1}$ and is retained if it is in the range $40 < m_1 < 120\GeV$.
A second $\ell^{+}\ell^{-}$ pair, denoted $\Z_{2}$, is required to have an invariant mass $12 < m_2 < 120\GeV$.
If more than one $\Z_{2}$ candidate satisfies all criteria, the pair of
leptons with the highest scalar $\pt$ sum is chosen.
The lepton $\pt$ selection is tightened with respect to the trigger by requiring at least one lepton to have $\pt > 20\GeV$, another one to have
$\pt > 10\GeV$, and any oppositely charged pair of leptons among the four selected to satisfy
$m_{\ell\ell} > 4\GeV$ regardless of flavor.
A $\cPZ$ boson decay into a lepton pair can be accompanied by final state radiation where
the radiated photon is associated to the corresponding lepton to form the $\cPZ$ boson candidate as
$\cPZ \to \ell^+\ell^-\gamma$~\cite{Chatrchyan:2013mxa}.

The electrons and muons that comprise the four-lepton candidate are checked for consistency with a reference vertex.
In the width analysis, this comparison is done with respect to the primary vertex of each event,
defined as the one passing the standard vertex requirements~\cite{Chatrchyan:2014fea} and having the largest $\sum \pt^2$ of all associated charged tracks.
The significance of the three-dimensional impact parameter (${\rm SIP}$) of each lepton, calculated from the track parameters and their uncertainties at the point of closest approach to this primary vertex, is required to be less than $4$~\cite{Chatrchyan:2013mxa}.
This requirement does not allow for a displaced vertex, so in order to constrain the lifetime of the \PH boson,
the reference of the comparison is switched to the vertex formed by the two leptons from the $\Z_1$ candidate.
The ${\rm SIP}$ of the two leptons from $\Z_1$ is required to be less than $4$, and that of the remaining two leptons is required to be less than 5.
An additional requirement $\chi^2_{4\ell} /\mathrm{dof}<6$ for the four-lepton vertex is applied to further suppress the $\Z+X$
background.
Both analyses also require the presence of the reconstructed proton-proton collision vertex in each event.
The combination of these requirements allows for the detection of a displaced \PH boson decay
while keeping the selection efficiencies similar between the two criteria.

After selection, the prompt-decay backgrounds originate from the $\Pq\Paq \to \Z\Z/\Z\gamma^*\to 4\ell$ and
$\Pg\Pg \to \Z\Z/\Z\gamma^*\to 4\ell$ processes together with $4\ell$ production with associated fermions, such as VBF
and associated $\V$ production. These backgrounds are evaluated from simulation following
Refs.~\cite{Chatrchyan:2013mxa,CMS-HIG-14-002}.
The $\Z+X$ background may include displaced vertices due to b-quark jets and is evaluated using the observed control samples as discussed in Ref.~\cite{Chatrchyan:2013mxa},
which employs the tight-to-loose lepton misidentification method.
While the misidentification rates are consistent between the two different vertex selection requirements, the overall number of selected
$\Z + X$ background events is about 15\% higher when using the vertex requirements of the lifetime measurement.
The number of prompt-decay signal and background events is about 2\% higher with these lifetime measurement requirements.

In the width analysis, the presence of jets is used as an indication of VBF or associated production
with an electroweak boson decaying hadronically, such as $\PW\PH$ or $\cPZ\PH$.
The CMS particle-flow (PF) algorithm~\cite{CMS-PAS-PFT-09-001,CMS-PAS-PFT-10-001,CMS-PAS-PFT-10-002,CMS-PAS-PFT-10-003},
which combines information from all subdetectors, is used to provide an event description in the form of reconstructed particle
candidates. The PF candidates are then used to build jets and lepton isolation quantities.
Jets are reconstructed using the anti-\kt clustering
algorithm~\cite{antikt} with a distance parameter of 0.5, as implemented in
the \textsc{FastJet} package~\cite{Cacciari:2011ma,Cacciari:fastjet2}. Jet energy corrections are applied as a
function of the jet $\pt$ and $\eta$~\cite{cmsJEC}.
An offset correction based on the jet area method is applied to subtract the energy contribution
not associated with the high-$\pt$ scattering such as electronic noise and pileup, the latter
of which results primarily from other pp collisions in the same bunch crossing~\cite{cmsJEC,Cacciari:2007fd,Cacciari:2008gn}.
Jets are only considered if they have $\pt>30\GeV$ and $\abs{\eta}<4.7$,
and if they are separated from the lepton candidates and identified final-state radiation photons.

Within the tracker acceptance, the jets are reconstructed with the
constraint that the charged particles are compatible with the primary
vertex.  In addition, jets arising from the primary
interaction are separated using a multivariate discriminator from those reconstructed due to energy deposits associated
with pileup interactions, particularly those from neutral particles not
associated with the primary vertex of the event. The discrimination is
based on the differences in the jet shapes, the relative multiplicity
of charged and neutral components, and the fraction of $\pt$ carried by the hardest components~\cite{jetIdPAS}.
In the width analysis, the events are split into two categories:
those with two or more selected jets (dijet category) and the remaining events (nondijet category).
When more than two jets are selected, the two jets with the highest $\pt$ are chosen for
further analysis.

The systematic uncertainties in the event selection are generally the same as those investigated in
Refs.~\cite{Chatrchyan:2013mxa,CMS-HIG-14-002,Khachatryan:2014kca,Khachatryan:2015cwa}.
Among the yield uncertainties, experimental systematic uncertainties are evaluated from data
for the lepton trigger efficiency and the combination of object reconstruction, identification, and isolation efficiencies.
Signal and background uncertainties after the lifetime analysis selection are found to be consistent with the width analysis selection.
Most of the signal normalization uncertainties are statistical in nature because the signal strength
is left unconstrained and because the systematic uncertainties affect only the relative
efficiency of $4\Pe$, $4\Pgm$, and $2\Pe2\Pgm$ reconstruction.
The overall predicted signal cross section is, therefore, not directly used in the analysis while
the theoretical uncertainties in the $4\ell$ background remain
unchanged compared to Refs.~\cite{Chatrchyan:2013mxa,CMS-HIG-14-002}.
The $\Z + X$ yield uncertainties are estimated to be 20\%, 40\%, and 25\%
for the $4\Pe$, $4\Pgm$, and $2\Pe2\Pgm$ decay channels, respectively, and also remain unchanged compared to Ref.~\cite{Chatrchyan:2013mxa}.

\section{Observables} \label{sec:observables}

Several observables, such as the four-lepton invariant mass, $\mell$, or the measured lifetime of each \PH boson candidate, $\VDlife$,
are used either as input to likelihood fits or to categorize events in this paper.
The full list of observables in each category is shown in Table~\ref{tab:kdlist}, and they are discussed in detail below.
The full kinematic information from each event is extracted using the MELA kinematic discriminants,
which make use of the correlation between either the two jets and the \PH boson
to identify the production mechanism, or the $\PH\to4\ell$ decay products to identify the decay kinematics.
These discriminants use either five, in the case of production, or seven, in the case of decay, mass and angular
input observables $\vec\Omega$~\cite{Gao:2010qx,Anderson:2013afp} to describe kinematics at LO in QCD\@.
The \pt of either the combined \PH boson and 2 jets system for the production discriminant
($\mathcal{D}_\text{jet}$)~\cite{Khachatryan:2015cwa} or the \PH boson itself for the decay discriminants ($\mathcal{D}^\text{ kin}$)~\cite{Chatrchyan:2012ufa}
is not included in the input observables in order to reduce associated uncertainties.

\begin{table*}[tbh]
\centering
\topcaption{
List of observables, $\vec{x}$, and categories of events used in the analyses of the \PH boson lifetime and width.
The $\mathcal{D}_\text{jet}< 0.5$ requirement is defined for $N_\text{jet}\ge 2$, but by convention this category also includes events with less than two selected jets, $N_\text{jet}< 2$.
}
\label{tab:kdlist}
\begin{scotch}{lccccc}
Category & Mass region & Criterion & \multicolumn{3}{c}{Observables $\vec{x}$} \\
\hline
Lifetime & $105.6 < \mell<140.6$\GeV & Any &  $\VDlife$ & $\mathcal{D}_\text{bkg}$  & \\
\hline
Width, on-shell dijet & $105.6 < \mell<140.6$\GeV & $N_\text{jet}\ge 2$ & $\mell$ & $\mathcal{D}^\text{kin}_\text{bkg}$ & $\mathcal{D}_\text{jet}$ \\
Width, on-shell nondijet & $105.6 < \mell<140.6$\GeV & $N_\text{jet}< 2$ & $\mell$ & $\mathcal{D}^\text{kin}_\text{bkg}$ & ${p}_\mathrm{T}$ \\
Width, off-shell dijet & $220 < \mell<1600$\GeV & $\mathcal{D}_\text{jet}\ge 0.5$ & $\mell$ & $\mathcal{D}_{\cPg\cPg}$ &  \\
Width, off-shell nondijet & $220 < \mell<1600$\GeV & $\mathcal{D}_\text{jet}< 0.5$ & $\mell$ & $\mathcal{D}_{\cPg\cPg}$ &  \\
\end{scotch}
\end{table*}

The discriminant sensitive to the VBF signal topology is calculated as
\begin{equation}
\label{eq:kd-vbfmela}
\mathcal{D}_\text{jet} =
\left[1+
\frac{ \mathcal{P}_{\PH\mathrm{JJ}} (\vec\Omega^{\PH+\mathrm{JJ}}, \mell) }
{\mathcal{P}_\mathrm{VBF}  (\vec\Omega^{\PH+\mathrm{JJ}}, \mell)  }\right]^{-1},
\end{equation}
where $\mathcal{P}_\mathrm{VMF}$ and $\mathcal{P}_{\PH \mathrm{JJ}}$ are probabilities obtained from the
\textsc{JHUGen} matrix elements for the VBF process and gluon fusion in association with two jets ($\PH+2\text{jets}$) within the
MELA framework~\cite{Khachatryan:2015cwa}.
This discriminant is equally efficient in separating VBF from either $\Pg\Pg\to\PH+2\text{jets}$ signal or $\Pg\Pg$
or $\qqbar\to 4\ell+2\text{jets}$ background because jet correlations in these processes are distinct from the VBF process.

\begin{figure}[tb!h]
\centering
\includegraphics[width=0.48\textwidth]{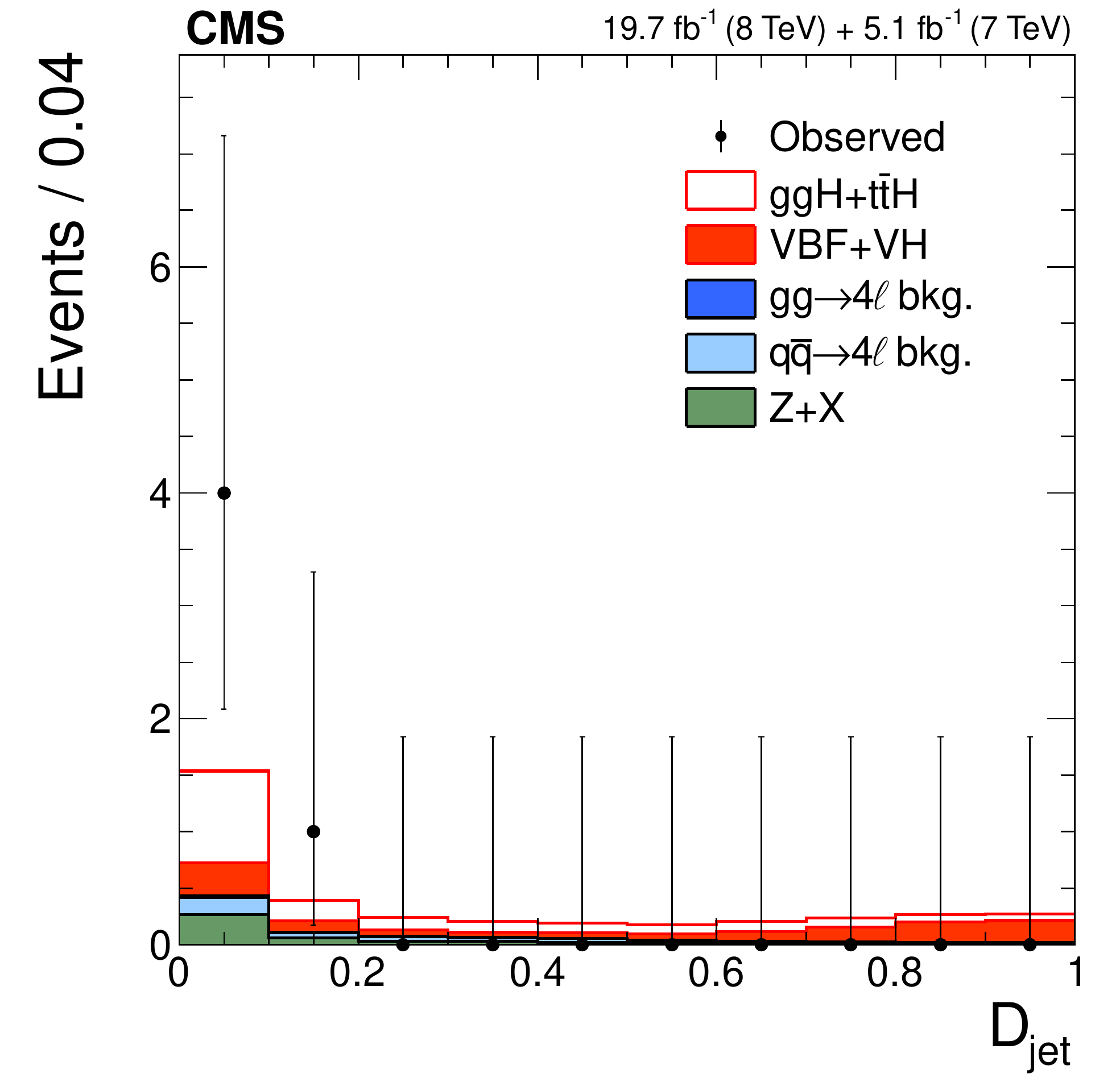}
\includegraphics[width=0.48\textwidth]{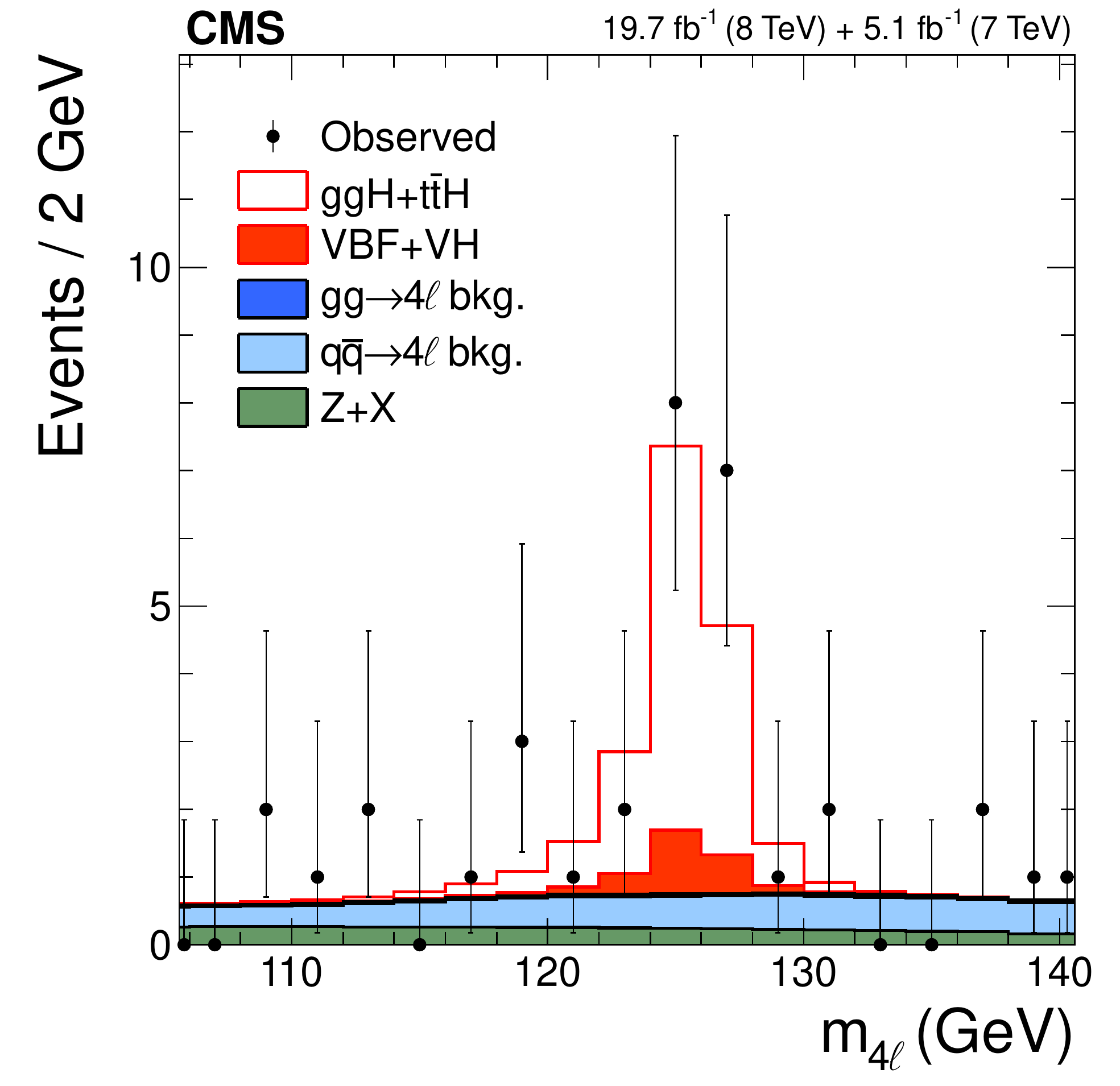}
\caption{
Distributions of the four-lepton invariant mass $\mathcal{D}_\text{jet}$ (\cmsLeft) and $\mell$ (\cmsRight) in the on-shell region
of the \PH boson width analysis.
The $\mathcal{D}_\text{jet}$ distributions show events in the dijet category with a requirement $120<\mell<130$\GeV.
The $\mell$ distributions combine the nondijet and dijet categories, the former with an additional
requirement $\mathcal{D}^\text{kin} _\text{bkg} > 0.5$ to suppress the dominant $\qqbar\to4\ell$ background.
The points with error bars represent the observed data,
and the histograms represent the expected contributions from the
SM backgrounds and the \PH boson signal. The contribution from the VBF and $\V\PH$ production is shown separately.
}
\label{fig:masslow}
\end{figure}

In the on-shell region, the $\mathcal{D}_\text{jet}$ discriminant is one of the width analysis observables used
in the dijet category. The $\mathcal{D}_\text{jet}$ distribution shown in Fig.~\ref{fig:masslow} (\cmsLeft) is used to distinguish gluon fusion, VBF, and $\V\PH$ production mechanisms in this category.
The \pt of the $4\ell$ system is used to distinguish the production mechanism of the remaining
on-shell events in the nondijet category.
In the off-shell region, the requirement $\mathcal{D}_\text{jet}\geq0.5$ is applied instead, keeping
nearly half of the VBF events and less than 4\% of all other processes, with
only a small dependence on $\mell$. Events that fail this requirement enter the nondijet category
in the off-shell region.
The different treatment of $\mathcal{D}_\text{jet}$ between the on-shell and off-shell regions
keeps the observables the same as in the previous width analysis~\cite{CMS-HIG-14-002}.

Uncertainties in modeling the jet distributions affect the separation of events between the
two dijet categories but do not affect the combined yield of either signal or background events.
For the on-shell dijet category, a 30\% normalization uncertainty is taken into account for the
$\Pg\Pg\to \PH + 2\text{jets}$ signal cross section while the uncertainty in the selection of two
or more jets from VBF production is 10\%.
The $\mathcal{D}_\text{jet}$ distribution uncertainties other than those for $\Z+X$ are estimated by comparing alternative MC generators and tunings,
where smaller effects from uncertainties due to jet energy scale and resolution are also included.

In the off-shell region, the uncertainties in the $\mathcal{D}_\text{jet}$ distribution imply uncertainties in the categorization requirement $\mathcal{D}_\text{jet}>0.5$.
To determine the uncertainty in the dijet selection, NLO QCD simulation
with \POWHEG is compared to the two LO generators
\textsc{phantom} and \textsc{JHUGen} for the VBF production, all with parton showering simulated with \PYTHIA.
For this comparison, $\V\PH$ production is omitted from the \textsc{phantom} simulation
since no events in association with electroweak boson production pass the $\mathcal{D}_\text{jet}\ge0.5$ requirement.
The efficiency of categorization for VBF-like events is stable within 5\%, and the main difference comes from the
uncertainty in the additional jet radiation after the hadronization of simulated events at LO or NLO in QCD using \PYTHIA.
A similar comparison of the signal production in gluon fusion is performed between the \POWHEG simulations at NLO in QCD with and without the \textsc{minlo} procedure for multijet simulation, and two LO
generators \MCFM and \textsc{gg2VV}. With proper matching of the hadronization scale for the LO generators in \PYTHIA~\cite{Chatrchyan:2014nva},
a good agreement within 15\% is found between all generators, with absolute dijet categorization efficiency of approximately~3\%. The $\mell$ dependence of the categorization efficiency is found to be similar between the different generators.

\begin{figure}[tb!hp]
\centering
\includegraphics[width=0.48\textwidth]{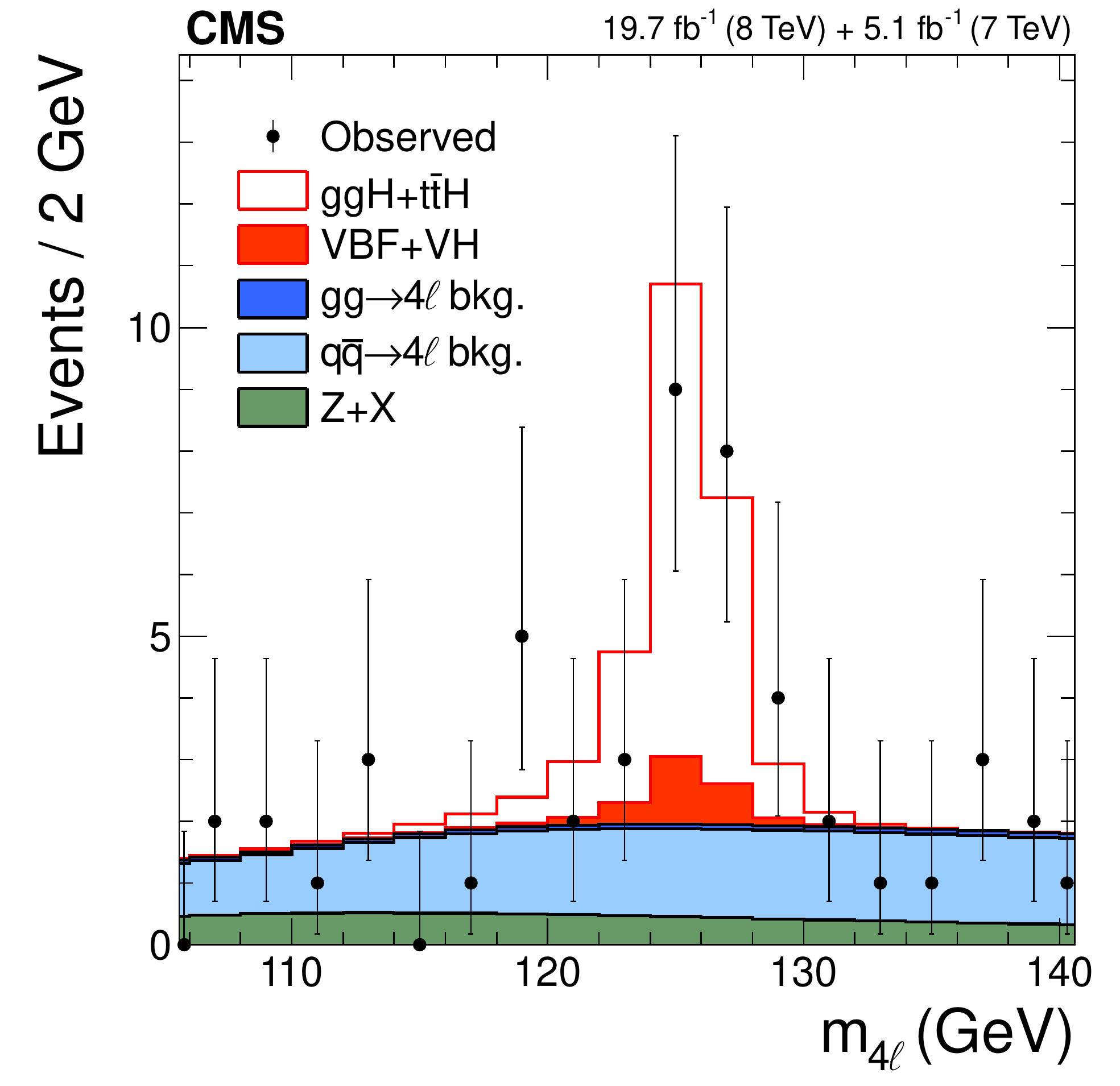}
\includegraphics[width=0.48\textwidth]{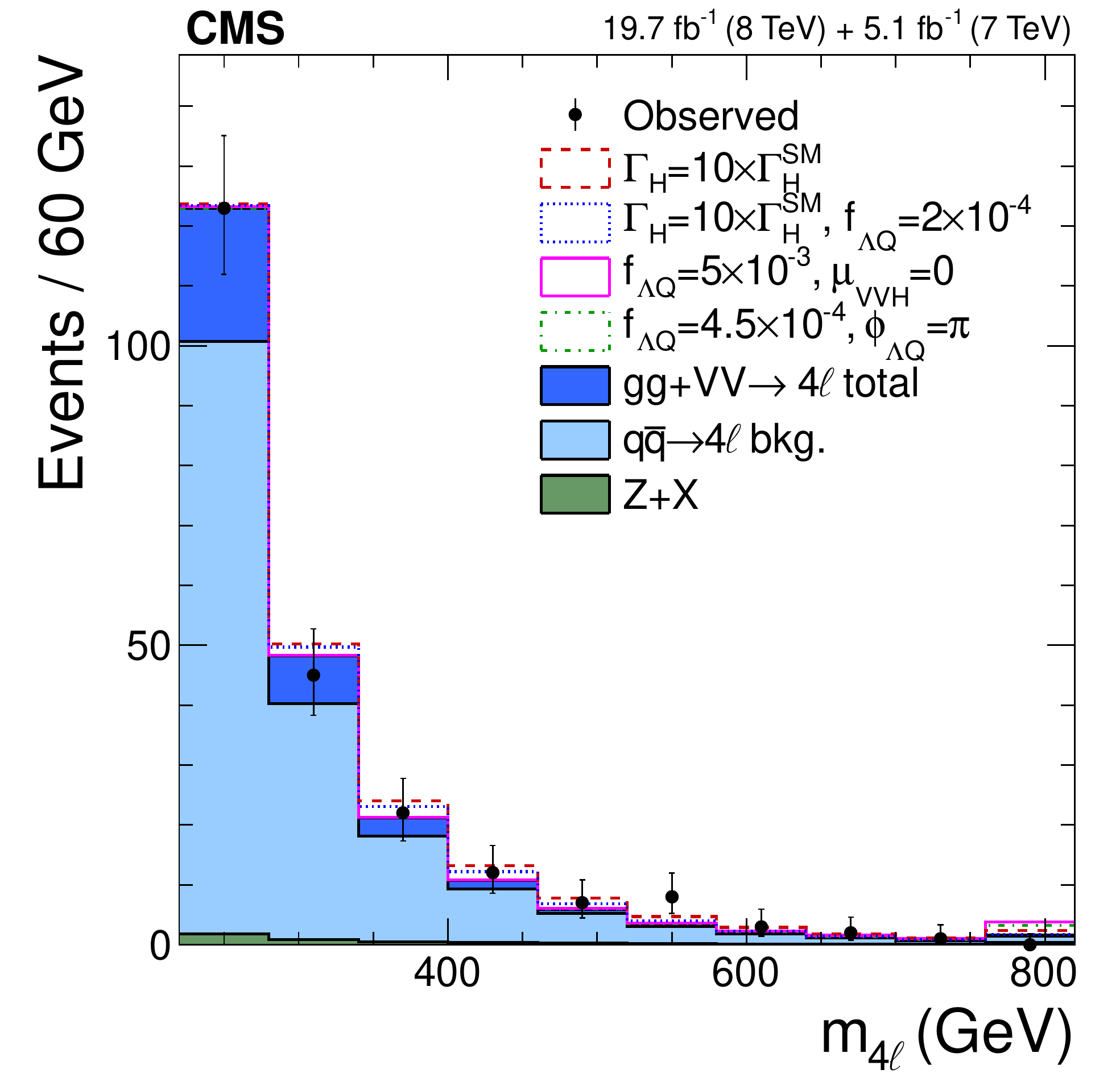}
\caption{
Distributions of the four-lepton invariant mass $\mell$ in the on-shell (\cmsLeft) and off-shell (\cmsRight) regions
of the \PH boson width analysis for all observed and expected events.
The points with error bars represent the observed data in both on-shell and off-shell region distributions.
The histograms for the on-shell region represent the expected contributions from the
SM backgrounds and the \PH boson signal with the contribution from the VBF and $\V\PH$ production shown separately.
The filled histograms for the off-shell region represent
the expected contributions from the
SM backgrounds and \PH boson signal, combining gluon fusion, VBF, and $\V\PH$ processes.
Alternative \PH boson width and coupling scenarios are shown as open histograms
with the assumption $\phiLQ=0$ unless specified otherwise,
and the overflow bin includes events up to $\mell=1600$\GeV.
}
\label{fig:massall}
\end{figure}

With the above uncertainties, the contributions of the signal, background, and their interference in the off-shell region for each category are obtained with the \textsc{phantom} generator for the VBF and
associated electroweak boson production, and
with the \MCFM generator for the gluon fusion production.
The dijet categorization efficiency as a function of $\mell$ is reweighted to the \POWHEG+ \textsc{minlo} prediction for gluon fusion signal contribution, and the same reweighting is used in the background and interference contributions.
For the $\qqbar \to \Z\Z$ background, the comparison of the NLO QCD simulation with \POWHEG with the two-jet inclusive
\MADGRAPH simulation leads to a 25\% uncertainty in the dijet categorization. Both dijet
categorization and its uncertainty have negligible $\mell$ dependence, and the dijet categorization efficiency is
around 0.6\%. An uncertainty of 100\% is assigned to the categorization of $\Z+X$ events, primarily due to statistical limitations in
the data-driven estimate. This uncertainty has a negligible contribution to the results since the contribution of $\Z+X$
is small in the total off-shell expected yield and negligible in the dijet category.

The discriminant sensitive to the $\Pg\Pg\to4\ell$ kinematics is calculated as
\begin{widetext}
\begin{equation}
\label{eq:kd-ggmela}
\mathcal{D}^\text{kin} =
\left[1+\frac{\alpha\, \mathcal{P}^{\Pq \Paq }_\text{bkg} (\vec\Omega^{\PH\to4\ell}, \mell)  }
{\mathcal{P}^{\cPg\cPg}_\text{sig}(\vec\Omega^{\PH\to4\ell}, \mell)
 +  \sqrt{\beta} \,  \mathcal{P}^{\cPg\cPg}_\text{int}(\vec\Omega^{\PH\to4\ell}, \mell)
 + \beta \, \mathcal{P}^{\cPg\cPg}_\text{bkg}(\vec\Omega^{\PH\to4\ell}, \mell)  } \right]^{-1},
\end{equation}
\end{widetext}
where the denominator contains the sum of the probability contributions from the signal
($\mathcal{P}^{\cPg\cPg}_\text{sig}$), the background ($\mathcal{P}^{\cPg\cPg}_\text{bkg}$),
and their interference ($\mathcal{P}^{\cPg\cPg}_\text{int}$) to the total $\cPg\cPg \to 4\ell$ process,
and the numerator includes the probability for the $\qqbar\to4\ell$ background process,
all calculated either with the \textsc{JHUGen} or \MCFM matrix elements within the MELA
framework~\cite{Chatrchyan:2013mxa,CMS-HIG-14-002,Khachatryan:2014kca}.
The two coefficients $\alpha$ and $\beta$ are tuned differently in the on-shell and off-shell
width analysis samples. Signal-background interference effects are
negligible in the on-shell region, so the kinematic discriminant is tuned to isolate signal from the dominant
background process with
$\mathcal{D}^\text{kin} _\text{bkg} = \mathcal{D}^\text{kin} (\alpha=1, \beta=0)$~\cite{Chatrchyan:2012ufa,Bolognesi:2012mm}.
In the off-shell region, the discriminant is tuned to isolate the full gluon fusion process, including the
interference term, for the ratio $\GHSM/\GH\sim\alpha=\beta=0.1$ close to the expected sensitivity of the analysis.
The discriminant is therefore labeled as
$\mathcal{D}_{\cPg\cPg}  = \mathcal{D}^\text{kin} (\alpha=\beta=0.1)$~\cite{CMS-HIG-14-002}.

\begin{figure}[tb!h]
\centering
\includegraphics[width=0.48\textwidth]{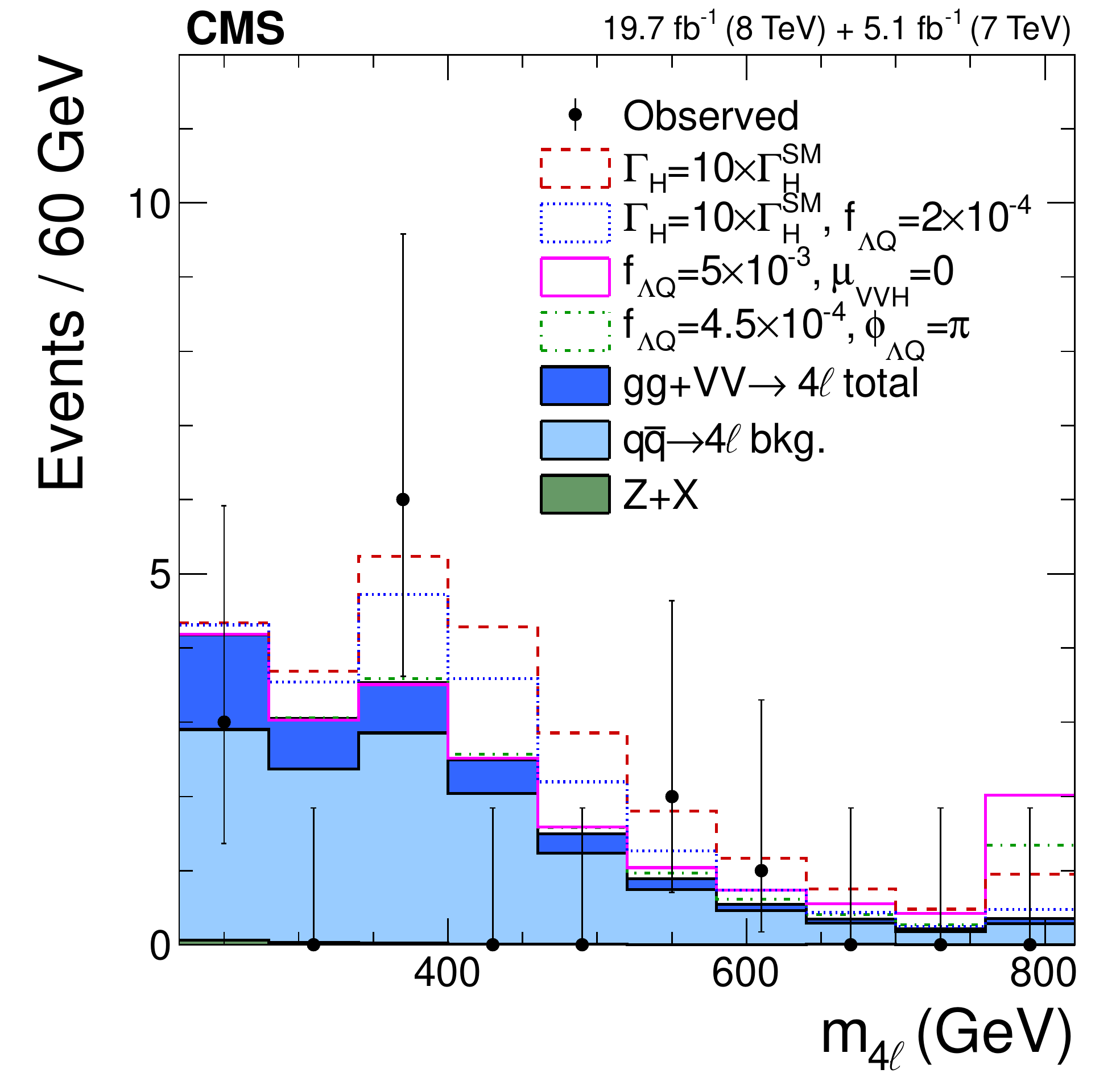}
\includegraphics[width=0.48\textwidth]{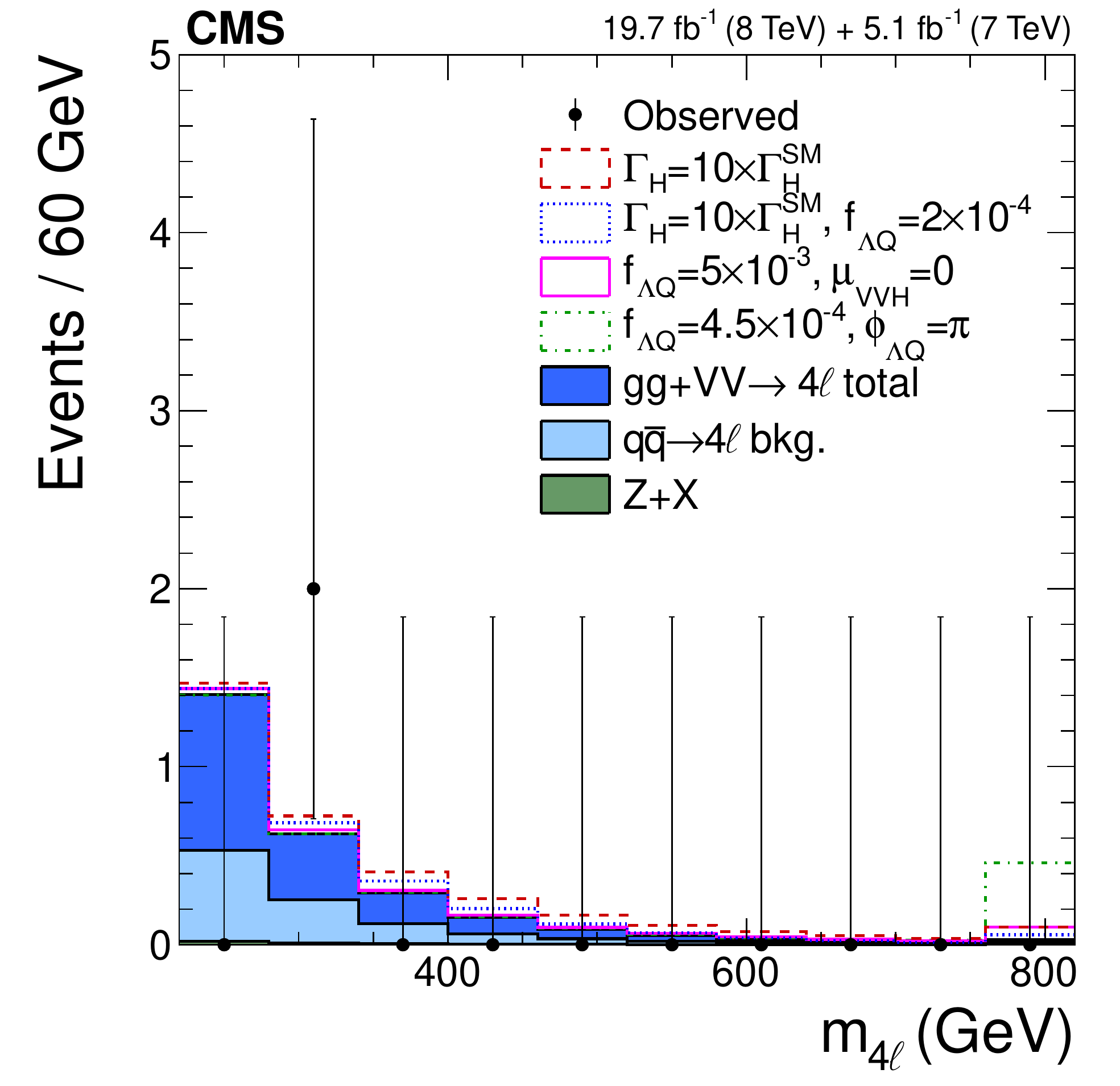}
\caption{
Distribution of the four-lepton invariant mass $\mell$ in the off-shell region in the nondijet (\cmsLeft)
and dijet (\cmsRight) categories. A requirement $\mathcal{D}_{\cPg\cPg} > 2/3$ is applied in the nondijet
category to suppress the dominant $\qqbar\to4\ell$ background.
The points with error bars represent the observed data, and the filled histograms represent
the expected contributions from the
SM backgrounds and \PH boson signal, combining gluon fusion, VBF, and $\V\PH$ processes.
Alternative \PH boson width and coupling scenarios are shown as open histograms.
The overflow bins include events up to $\mell=1600$\GeV, and $\phiLQ=0$ is assumed where it is unspecified.
}
\label{fig:massfull}
\end{figure}

\begin{figure}[tb!h]
\centering
\includegraphics[width=0.48\textwidth]{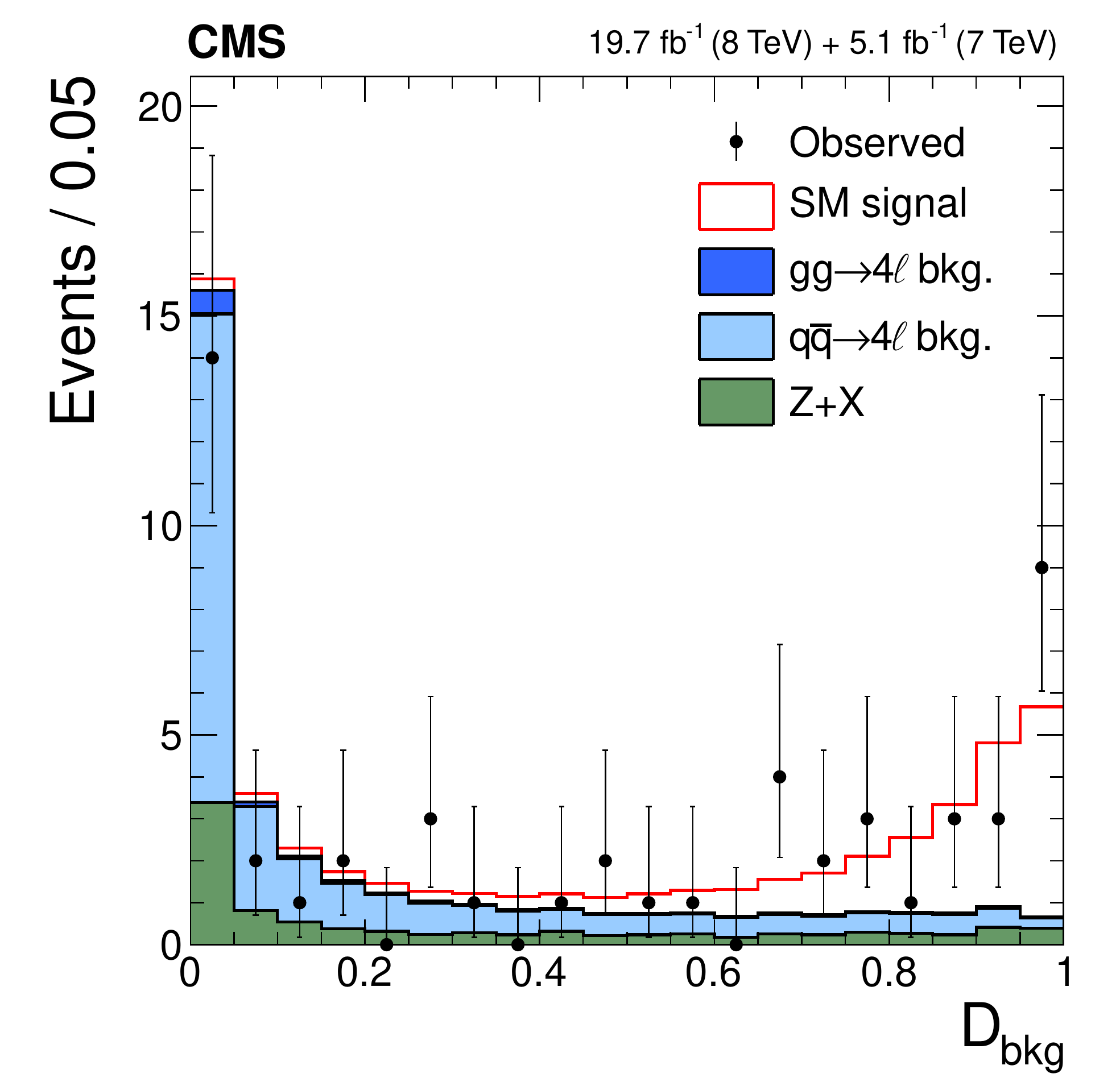}
\includegraphics[width=0.48\textwidth]{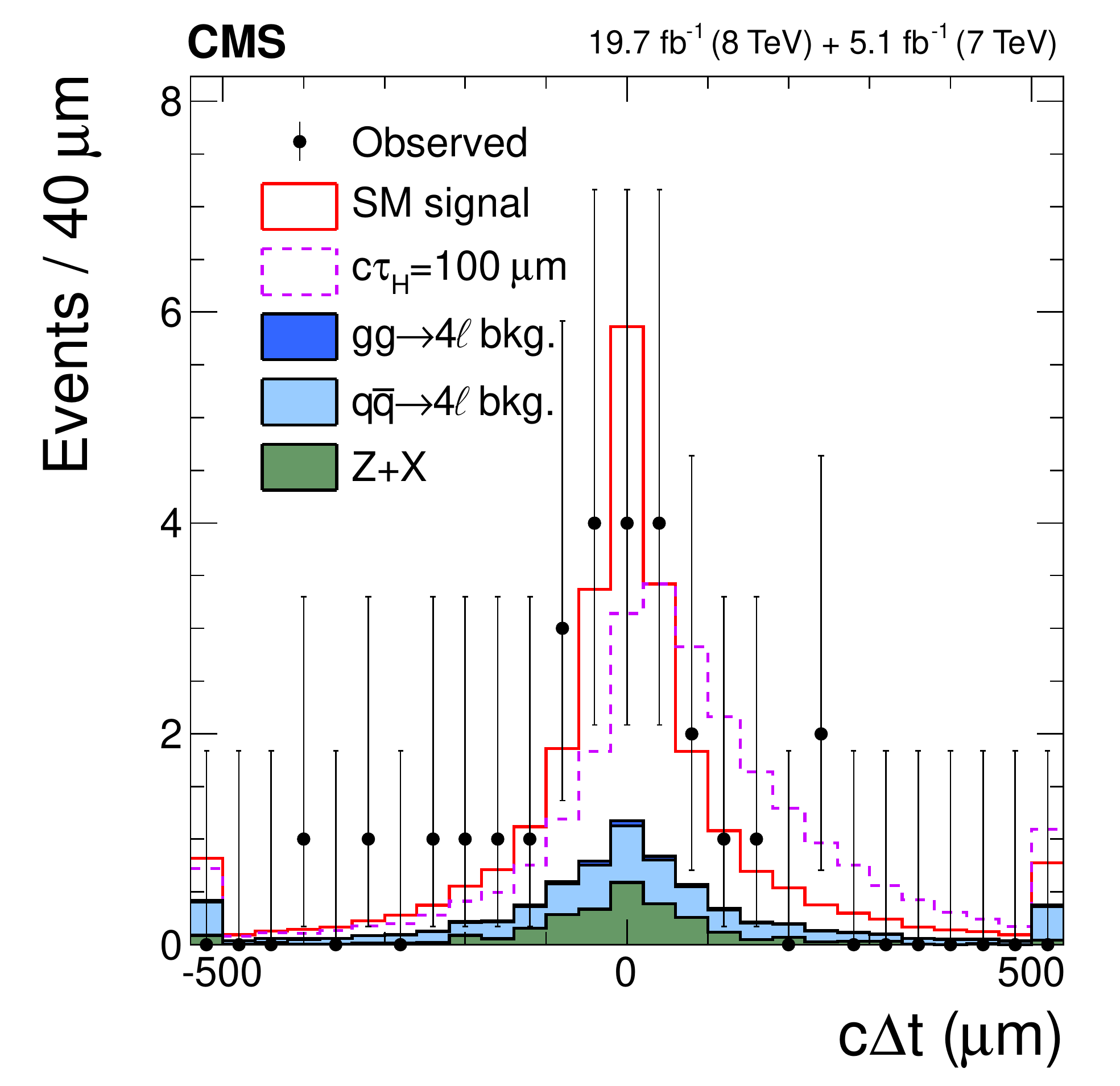}
\caption{
Distributions of $\mathcal{D}_\text{bkg}$ (\cmsLeft) and $c\VDlife$ (\cmsRight) in the lifetime analysis with
$\mathcal{D}_\text{bkg} > 0.5$ required for the latter to suppress the background.
The points with error bars represent the observed data,
and the filled histograms stacked on top of each other
represent the expected contributions from the SM backgrounds.
Stacked on the total background contribution,
the open histograms show
the combination of all production mechanisms expected in the SM
for the \PH boson signal
with either the SM lifetime or $c\tau_\PH=100\mum$.
Each signal contribution in the different open histograms
are the same as the total number of events
expected from the combination of all production mechanisms in the SM.
All signal distributions are shown with the total number of events expected in the SM.
The first and last bins of the $c\VDlife$ distributions include all events beyond $\abs{c\VDlife}>500\mum$.
}
\label{fig:discriminants}

\end{figure}

Apart from the above kinematic discriminants and \pt, the width analysis employs the four-lepton invariant mass $\mell$
as the main observable, which provides signal and background separation in the on-shell region and
which is sensitive to the $\GH$ values and anomalous couplings in the off-shell region.
The $\mell$ distributions are illustrated in
Fig.~\ref{fig:massall} for the on-shell and off-shell regions without any kinematic requirements,
Fig.~\ref{fig:masslow} (\cmsRight) for the on-shell region with the requirement $\mathcal{D}^\text{kin} _\text{bkg} > 0.5$,
and Fig. ~\ref{fig:massfull} for the two event categories in the off-shell region, with the requirement $\mathcal{D}_{\cPg\cPg} > 2/3$ on the nondijet category.
The requirements on the kinematic discriminants $\mathcal{D}^\text{kin} _\text{bkg}$ or $\mathcal{D}_{\cPg\cPg}$
suppress the relative contribution of background in the illustration of event distributions.
In the lifetime analysis,
the $\mell$ and $\mathcal{D}^\text{kin} _\text{bkg}$ observables are combined into one, called $\mathcal{D}_\text{bkg}$~\cite{Chatrchyan:2012jja,Anderson:2013afp,Khachatryan:2014kca}, in order to reduce the number of observables.
It is constructed by multiplying the matrix element probability ratio in Eq.~(\ref{eq:kd-ggmela}) by the ratio of probabilities
for $\mell$ from the nonresonant $\qqbar \to 4\ell$ process and the resonant production
$\Pg\Pg\to \PH\to 4\ell$ for the measured $m_\PH=125.6\GeV$.
The $\mathcal{D}_\text{bkg}$ distribution in the lifetime analysis is shown in Fig.~\ref{fig:discriminants}.
To account for the lepton momentum scale and resolution uncertainty in the $\mell$ or
$\mathcal{D}_\text{bkg}$ distributions, alternative signal distributions are taken from the variations
of both of these contributions.

\begin{figure}[thb]
\centering
\includegraphics[width=0.48\textwidth]{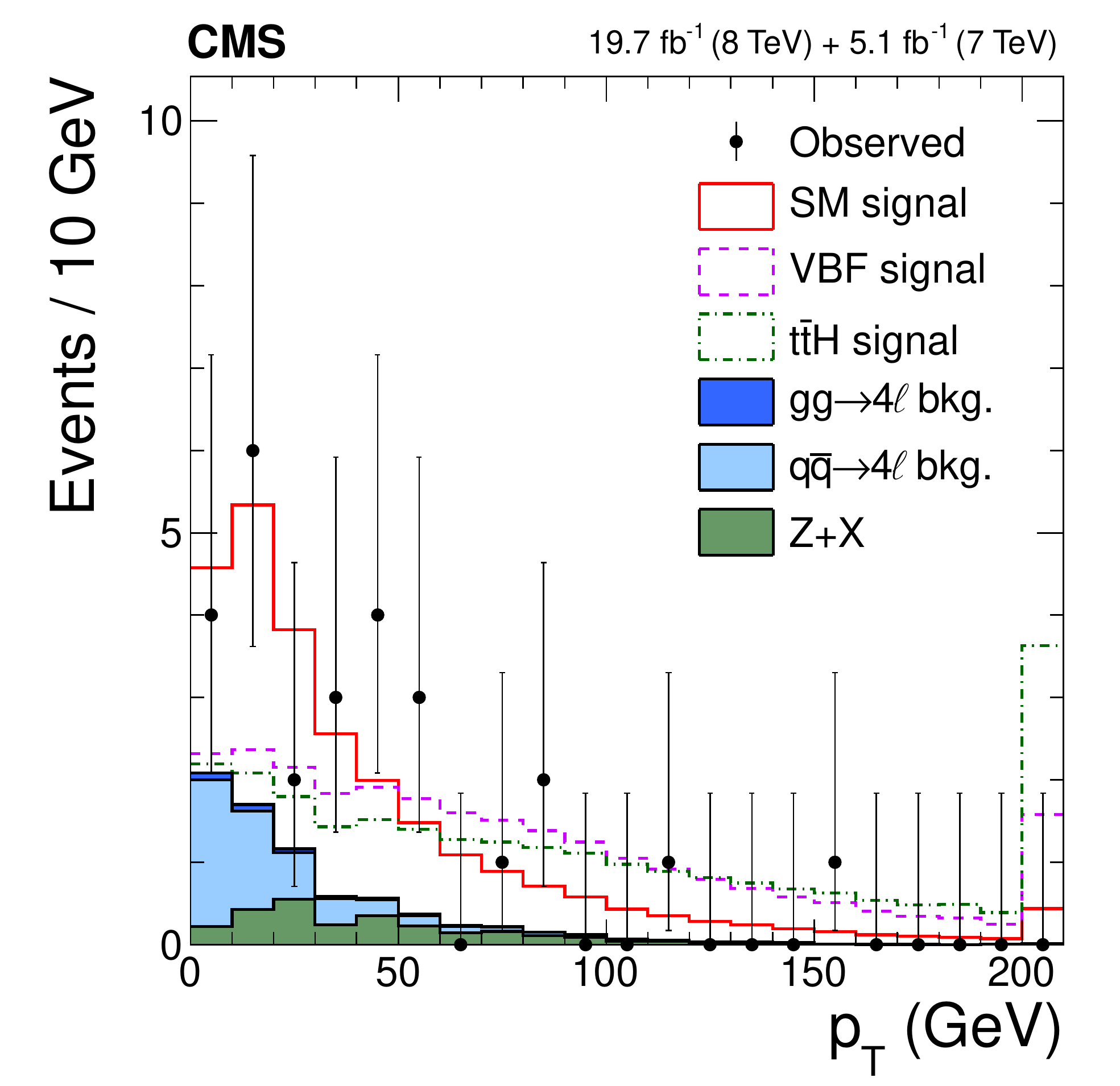}
\caption{
Distributions of the four-lepton \pt with the selection used in the lifetime analysis and
the requirement $\mathcal{D}_\text{bkg} > 0.5$ to suppress the backgrounds.
The points with error bars represent the observed data,
and the filled histograms stacked on top of each other
represent the expected contributions from the SM backgrounds.
Stacked on the total background contribution, the open histograms
show the \PH boson signal either with
the combination of all production mechanisms expected in the SM,
or for the VBF or $\ttbar\PH$ production mechanisms.
Each signal contribution in the different open histograms
is normalized to the total number of events
expected from the combination of all production mechanisms in the SM.
The overflow bin includes all events beyond $\pt>200$\GeV.
}
\label{fig:HpTinclusive}
\end{figure}

The lifetime analysis makes use of the observable $\VDlife$ calculated following Eq.~(\ref{eq:tau_define}). The reference point for \PH boson production vertex is taken to be the beam spot,
which is the pp collision point determined by fitting charged-particle tracks from events in multiple collisions,
and the value of  $\Delta\vec{r}_\mathrm{T}$ is calculated
as the displacement from the beam spot to the $4\ell$ vertex in the plane transverse to the beam axis.
An alternative calculation of $\VDlife$ has also been considered using the primary vertex of each event
instead of the beam spot, but the different associated particles in the \PH boson production and their multiplicity
would introduce additional model dependence in the primary vertex resolution.

The $\VDlife$ value is non-negative and follows the exponential decay distribution if it is known perfectly for each event.
However, resolution effects arising mostly from limited precision of the $\Delta\vec{r}_\mathrm{T}$ measurement
allow negative $\VDlife$ values. This feature allows for an effective self-calibration of the resolution from the data.
Symmetric broadening of the $\VDlife$ distribution indicates resolution effects while positive skew
indicates sizable signal lifetime. Figure~\ref{fig:discriminants} displays the $\VDlife$ distributions.
The resolution in $\VDlife$ also depends on the \pt spectrum of the produced \PH boson, which differs among the production mechanisms, and this dependence is accounted for in the fit procedure
as described in detail in Sec.~\ref{sec:lifetime}.
The distributions of $\VDlife$ and \pt are shown in Figs.~\ref{fig:discriminants} and~\ref{fig:HpTinclusive}, respectively. Since the discriminant $\mathcal{D}_\text{bkg}$ is optimal for signal separation in the on-shell region, a requirement $\mathcal{D}_\text{bkg}>0.5$ is applied to reduce the background when showing these distributions.

Uncertainties in the $\VDlife$ distribution for the signal and the prompt background are obtained
from a comparison of the expected and
observed distributions in the $\mell$ sidebands, $70< \mell<105.6$\GeV and $170< \mell<800$\GeV.
These uncertainties obtained from this comparison correspond to varying the $\VDlife$ resolution by
$+17$/$-15$\%, $+14$/$-12$\%, and $+20$/$-17$\% for the $4\Pe$, $4\Pgm$, and $2\Pe2\Pgm$ final states, respectively.
The $\Z+X$ parametrization is obtained from the control region in the analysis mass range
$105.6< \mell < 140.6$\GeV, and its alternative parametrization
obtained from the control region events in the mass range $140.6< \mell < 170$\GeV
reflects the uncertainties in the data-driven estimate.
A cross-check of the $\VDlife$ distributions is also performed with the $3\ell$ control samples
enriched in $\PW\Z$ prompt decay, and the distributions are found to be consistent with simulation.

\section{Constraints on the lifetime}  \label{sec:lifetime}

The \PH boson lifetime analysis is based on two observables $\vec{x}=(\VDlife, \mathcal{D}_\text{bkg})$,
which allow the measurement of the average signal lifetime $\tau_\PH$ and
the discrimination of the \PH boson signal from background using a simultaneous likelihood fit.
The extended likelihood function is defined for $N_\text{ev}$ candidate events as
\ifthenelse{\boolean{cms@external}}{
\begin{multline}
\mathcal{L} =  \exp\Bigl( - n_\text{sig}-\sum_k n_\text{bkg}^{( k )}  \Bigr)
\prod_i^{N_\text{ev}} \Bigl( n_\text{sig} \,\mathcal{P}_\text{sig}(\vec{x}_{i};\vec{\xi})\\
+\sum_k n_\text{bkg}^{( k )} \,\mathcal{P}_\text{bkg}^{\left( k \right)} (\vec{x}_{i};\vec{\zeta})
\Bigr),
\label{eq:likelihood}
\end{multline}
}{
\begin{equation}
\mathcal{L} =  \exp\left( - n_\text{sig}-\sum_k n_\text{bkg}^{( k )}  \right)
\prod_i^{N_\text{ev}} \left( n_\text{sig} \,\mathcal{P}_\text{sig}(\vec{x}_{i};\vec{\xi})
+\sum_k n_\text{bkg}^{( k)} \,\mathcal{P}_\text{bkg}^{\left( k \right)} (\vec{x}_{i};\vec{\zeta})
\right),
\label{eq:likelihood}
\end{equation}
}
where $n_\text{sig}$ is the number of signal events and $n_\text{bkg}^k$ is the number of
background events of type~$k$ ($\cPg\cPg \to 4\ell$, $\Pq \Paq \to 4\ell$, $\Z+X$).
The probability density functions $\mathcal{P}_\text{sig}$ for signal, and  $\mathcal{P}_\text{bkg}^k$
for each background process $k$ are described as histograms (templates).
The likelihood parametrization is constructed independently in each of the
$4\Pe$, $4\Pgm$, or $2\Pe2\Pgm$ final states, and for 7 and 8\TeV pp collision energy.
The parameters $\vec{\xi}$ for the signal and $\vec{\zeta}$ for the background processes
include parametrization uncertainties, and $\vec{\xi}$ also includes $\tau_\PH$ as the parameter of interest.
The likelihood in Eq.~(\ref{eq:likelihood}) is maximized with respect to the parameters $n_\text{sig}$, $n_\text{bkg}^k$, $\vec{\xi}$ and $\vec{\zeta}$, which constitute the nuisance parameters and the parameter of interest. The nuisance parameters are either constrained within the associated uncertainties or left unconstrained in the fit.

The kinematics of the four-lepton decay, affecting $\mathcal{D}_\text{bkg}$,
and the four-lepton vertex position and resolution, affecting $\VDlife$, are found to be independent.
Therefore, the two-dimensional probability distributions of $\mathcal{P}(\VDlife, \mathcal{D}_\text{bkg})$
are constructed as the product of two one-dimensional distributions.
In the case of the signal probability, the \VDlife\ templates are conditional on the parameter of interest $\tau_\PH$.
The signal \VDlife\ parametrization is obtained for the range $0\leq c\tau_\PH \leq 1000\,\mu$m
by reweighting the simulation available for the gluon fusion process at $c\tau_\PH=0,$ 100, 500, and 1000\mum
to $c\tau_\PH$ values in steps of 10\mum and interpolating linearly for any intermediate value.

The $\VDlife$ parametrization for all SM \PH boson production mechanisms
(gluon fusion, VBF, $\PW\PH$, $\Z\PH$, and $\ttH$) is obtained by reweighting gluon fusion
production events as a function of \pt at each of the $\tau_\PH$ values.
This procedure reproduces $\VDlife$ resolution effects predicted from the simulation
for prompt signal (\ie $\tau_\PH=0$) and is, therefore, valid for nonzero lifetime. As shown
in Fig.~\ref{fig:HpTinclusive}, the gluon fusion
production mechanism has the softest \pt spectrum while $\ttH$ production yields the hardest
\pt, and the distribution of $\VDlife$ is thus wider
in gluon fusion and narrower in $\ttH$ production, with other production mechanisms in between.
Gluon fusion production and $\ttH$ distributions, with their respective
yields scaled to the total SM production cross section, are therefore taken as the two extreme variations
while the nominal $\VDlife$ distribution is parametrized with the SM combination
of the different production mechanisms.
The $\VDlife$ distribution used in the likelihood is varied from the nominal prediction between these two extremes with a continuous production parameter included in $\vec{\xi}$ in Eq.~(\ref{eq:likelihood}).
Any other production mechanism or a mixture can be
described with this parametrization, and the values of the production parameter
corresponding to the \pt spectrum of either pure VBF, $\PW\PH$, $\Z\PH$, or $\ttH$
mechanisms are excluded at more than 95\% \CL from a fit to data.
This information is consistent with the observed \pt spectrum in Fig.~\ref{fig:HpTinclusive}.

\begin{figure}[tbhp]
\centering
\includegraphics[width=\cmsFigWidth]{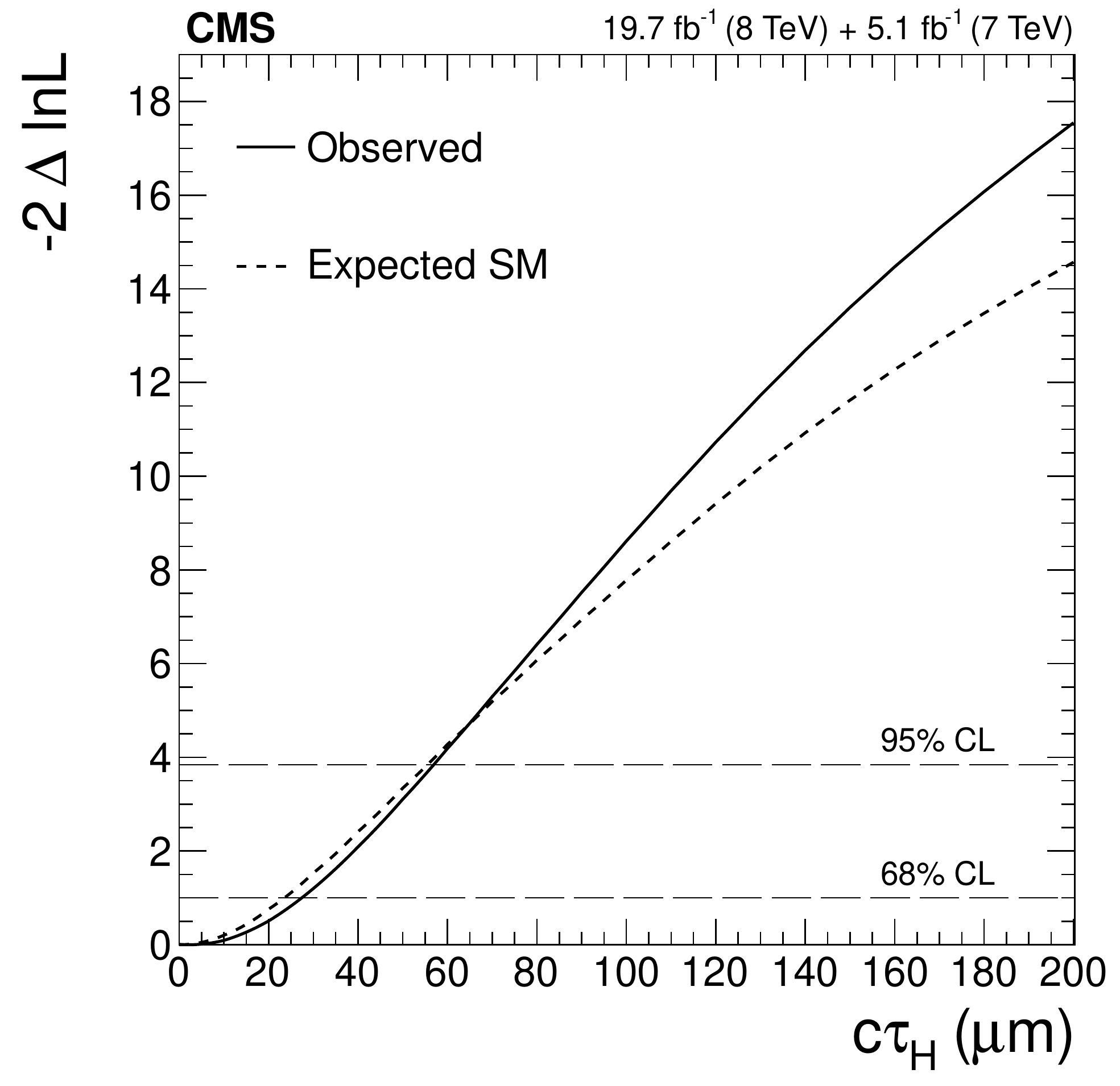}
\caption{
Observed (solid) and expected (dashed) distributions of $-2\ln(\mathcal{L}/\mathcal{L}_\text{max})$
as a function of the \PH boson average lifetime $c\tau_\PH$.
}
\label{fig:lifetime_nocat}

\end{figure}

While the $\VDlife$ and $\mathcal{D}_\text{bkg}$ parametrizations are obtained for the SM couplings
in the $\PH\to \Z\Z\to 4\ell$ decay, and for \pt spectra
as in SM-like production mechanisms (gluon fusion, VBF, $\PW\PH$, $\Z\PH$, and $\ttH$),
the analysis has little dependence on anomalous couplings in either the production or the decay of the \PH boson.
It has already been established~\cite{Khachatryan:2014kca} that the kinematics of the $\PH\to 4\ell$ decay
are consistent with the kinematics of the SM \PH boson decay and inconsistent with a wide range of exotic models.
The $\VDlife$ and $\mathcal{D}_\text{bkg}$ distributions have little variation within the
allowed range of exotic couplings in the $\PH\to 4\ell$ decay.
The expected $\tau_\PH$ constraint remains stable within 10\% when the simulation for those exotic models is tested instead of the simulation with SM couplings.
Anomalous couplings in production are found to have a substantial effect on the
\pt spectrum, typically making the spectrum harder in the VBF, $\PW\PH$, $\Z\PH$, and $\ttH$
production mechanisms. Extreme variations in the \pt spectrum, however, are already excluded by the data,
and \pt variations allowed by the data are reflected in the $\VDlife$ parametrization with the
parameter describing the production mechanisms.

Figure~\ref{fig:lifetime_nocat} shows the likelihood distribution as a function of $c\tau_\PH$.
The allowed 68\% and 95\% \CL intervals are defined using the respective profile likelihood function values $-2\ln(\mathcal{L}/\mathcal{L}_\text{max}) = 1.00$ and 3.84 for which exact coverage is expected in the asymptotic limit~\cite{wilks1938}. The approximate coverage
has been tested with the generated samples at different $c\tau_\PH$ values, and the quoted results have been found to be conservative.
The observed (expected) average lifetime is
$c\tau_\PH=2^{+25}_{-2}$ ($0^{+24}_{-0}$)\mum
($\tau_\PH=10^{+80}_{-10}$\unit{fs} for the observation and $\tau_\PH=0^{+80}_{-0}$\unit{fs} for the expectation),
and the allowed region at the 95\% \CL is $c\tau_\PH<57\,(56)\mum$
($\tau_\PH<190$\unit{fs} for both the observation and the expectation).
The observed number of signal events remains consistent with Ref.~\cite{Chatrchyan:2013mxa}.
The observed (expected) upper limit on the average lifetime at 95\% \CL corresponds
through Eq.~(\ref{eq:tauHgammaH}) to
the lower limit on the \PH boson width $\GH > 3.5 \times 10^{-9}$\MeV ($\GH > 3.6 \times 10^{-9}$\MeV)
regardless of the value of $\fLQ$.

\section{Constraints on the width}  \label{sec:width}

The \PH boson width $\GH$ and the effective fraction $\fLQ$ for the $\Lambda_Q$ anomalous coupling are measured
in an unbinned maximum likelihood fit of a signal-plus-background model
following Eq.~(\ref{eq:likelihood}). In addition to the event categories already
defined in the lifetime analysis for the final states and pp collision energy, events are also split into dijet and nondijet categories, and into on-shell and off-shell regions. In the on-shell region, a three-dimensional distribution
of $\vec{x}= (\mell, \mathcal{D}_\text{bkg}^\text{kin}, \pt$ or $\mathcal{D}_\text{jet})$
is analyzed, following the methodology described in Ref.~\cite{Chatrchyan:2013mxa}.
In the off-shell region, a two-dimensional distribution $\vec{x}=(\mell, \mathcal{D}_{\cPg\cPg})$
is analyzed following the methodology described in Ref.~\cite{CMS-HIG-14-002}
with the events split into the two dijet categories defined in Table~\ref{tab:kdlist}.

The probability distribution functions are built using the full detector simulation or data control
regions and are defined for both the signal ($\mathcal{P}_\text{sig}$) and the background
($\mathcal{P}_\text{bkg}$) contributions as well as their interference ($\mathcal{P}_{\text{int}}$), as a function of
the observables $\vec{x}$ discussed above. Several production mechanisms such as gluon fusion (gg), VBF, $\PW\PH$ and $\Z\PH$ ($\V\PH$) are considered for the signal.
The total probability distribution function for the off-shell region is written as
\ifthenelse{\boolean{cms@external}}{
\begin{multline}
\mathcal{P}_\text{tot}^\text{off-shell}(\vec{x}; \GH, \fLQ)  =
\Biggl[
\mu_{\Pg\Pg\PH}\,\frac{\GH}{\Gamma_0} \, \mathcal{P}^{gg}_\text{sig}(\vec{x}; \fLQ)\\
+  \sqrt{\mu_{\Pg\Pg\PH}\,\frac{\GH}{\Gamma_0} } \, \mathcal{P}^{gg}_\text{int}(\vec{x}; \fLQ)
+  \mathcal{P}^{gg}_\text{bkg}(\vec{x})
\Biggr] \\
  +
\Biggl[
\mu_{\V\V\PH}\,\frac{\GH}{\Gamma_0} \, \mathcal{P}^{\V\V}_\text{sig}(\vec{x}; \fLQ)
+  \sqrt{\mu_{\V\V\PH} \,\frac{\GH}{\Gamma_0} } \, \mathcal{P}^{\V\V}_\text{int}(\vec{x}; \fLQ)
\\+ \mathcal{P}^{\V\V}_\text{bkg}(\vec{x})
\Biggr]
 + \mathcal{P}^{\Pq\Paq}_\text{bkg}(\vec{x}) +
\mathcal{P}^{\Z X}_\text{bkg}(\vec{x}),
\label{eq:offshell-prob}
\end{multline}
}{
\begin{multline}
\mathcal{P}_\text{tot}^\text{off-shell}(\vec{x}; \GH, \fLQ)  =
\left[
\mu_{\Pg\Pg\PH}\,\frac{\GH}{\Gamma_0} \, \mathcal{P}^{gg}_\text{sig}(\vec{x}; \fLQ)
+  \sqrt{\mu_{\Pg\Pg\PH}\,\frac{\GH}{\Gamma_0} } \, \mathcal{P}^{gg}_\text{int}(\vec{x}; \fLQ)
+  \mathcal{P}^{gg}_\text{bkg}(\vec{x})
\right] \\
  +
\left[
\mu_{\V\V\PH}\,\frac{\GH}{\Gamma_0} \, \mathcal{P}^{\V\V}_\text{sig}(\vec{x}; \fLQ)
+  \sqrt{\mu_{\V\V\PH} \,\frac{\GH}{\Gamma_0} } \, \mathcal{P}^{\V\V}_\text{int}(\vec{x}; \fLQ)
+ \mathcal{P}^{\V\V}_\text{bkg}(\vec{x})
\right]
\\
 + \mathcal{P}^{\Pq\Paq}_\text{bkg}(\vec{x}) +
\mathcal{P}^{\Z X}_\text{bkg}(\vec{x}),
\label{eq:offshell-prob}
\end{multline}
}
where $\Gamma_0$ is a reference value used in simulation and $\V\V$ stands for a combination
of VBF and associated electroweak boson production taken together.
Under the assumption $\phiLQ=0$ or $\pi$, any contribution
to the $\PH\V\V$ scattering amplitude in Eq.~(\ref{eq:fullampl-formfact-spin0})
from the $a_1$ term
is proportional to $\sqrt{1-\fLQ }$
while that
from the $\Lambda_{Q}$ term
is proportional to $\sqrt{\fLQ }  \, \cos \left( \phiLQ \right)$.
The dependence on \fLQ\ in Eq.~(\ref{eq:offshell-prob}) can thus be parametrized with the factor
\begin{equation}
\left(
\sqrt{1-\fLQ } - \sqrt{\fLQ }  \, \cos \left( \phiLQ \right) \, \frac{\mell^2}{m_{\PH}^2}
\right)^N,
\label{eq:offshell-factor}
\end{equation}
where the power $N$ depends on the power of the $\PH\V\V$ couplings.
The couplings appear twice in the VBF and $\V\PH$ cases, in both production and decay,
so the power of the factor is twice as large.
Thus, for gluon fusion, $N=1$ for the interference component ($\mathcal{P}^{gg}_\text{int}$) and $N=2$ for the signal ($\mathcal{P}^{gg}_\text{sig}$); for $\mathrm{VBF}$ and $\V\PH$, $N=2$ ($\mathcal{P}^{\V\V}_\text{int}$) and $4$ ($\mathcal{P}^{\V\V}_\text{sig}$), respectively.
Both $\PH\Z\Z$ and $\PH\PW\PW$ couplings contribute to the VBF and $\V\PH$ production couplings,
and this analysis assumes the same $\Lambda_Q$ would contribute to the $\PH\Z\Z$ and $\PH\PW\PW$
couplings in Eq.~(\ref{eq:fullampl-formfact-spin0}).
The effective fraction $\fLQ$ is therefore the same for the $\PH\Z\Z$ and $\PH\PW\PW$ amplitudes.

In the on-shell region, the parametrization includes the small contribution of the $\ttH$
production mechanism, which is related to the gluon fusion production. The total probability
distribution function for the on-shell region is
\ifthenelse{\boolean{cms@external}}{
\begin{multline}
 \mathcal{P}_\text{tot}^\text{on-shell}(\vec{x})  =
\mu_{\Pg\Pg\PH}\, \mathcal{P}^{\Pg\Pg+\ttbar\PH}_\text{sig}(\vec{x})
+ \mu_{\V\V\PH}  \,  \mathcal{P}^{\V\V}_\text{sig}(\vec{x})
 \\+ \mathcal{P}^{\Pq\Paq}_\text{bkg}(\vec{x}) + \mathcal{P}^{\Pg\Pg}_\text{bkg}(\vec{x}) +  \mathcal{P}^{\Z X}_\text{bkg}(\vec{x}).
\label{eq:pdf-prob-onshell}
\end{multline}
}{
\begin{equation}
 \mathcal{P}_\text{tot}^\text{on-shell}(\vec{x})  =
\mu_{\Pg\Pg\PH}\, \mathcal{P}^{\Pg\Pg+\ttbar\PH}_\text{sig}(\vec{x})
+ \mu_{\V\V\PH}  \,  \mathcal{P}^{\V\V}_\text{sig}(\vec{x})
 + \mathcal{P}^{\Pq\Paq}_\text{bkg}(\vec{x}) + \mathcal{P}^{\Pg\Pg}_\text{bkg}(\vec{x}) +  \mathcal{P}^{\Z X}_\text{bkg}(\vec{x}).
\label{eq:pdf-prob-onshell}
\end{equation}
}
The normalization of the signal and background distributions is incorporated in the probability functions $\mathcal{P}$
in Eqs.~(\ref{eq:offshell-prob})~and~(\ref{eq:pdf-prob-onshell}), but the overall signal yield is left
unconstrained with the independent signal strength parameters $\mu_{\Pg\Pg\PH}$ and $\mu_{\V\V\PH}$,
corresponding to the \PH production mechanisms through coupling to either fermions or weak vector bosons, respectively.
The observed $\mu_{\Pg\Pg\PH}$ and $\mu_{\V\V\PH}$ values are found to be consistent with those obtained
in Refs.~\cite{Chatrchyan:2013mxa,CMS-HIG-14-002}.

\begin{figure}[tbh]
\centering
\includegraphics[width=\cmsFigWidth]{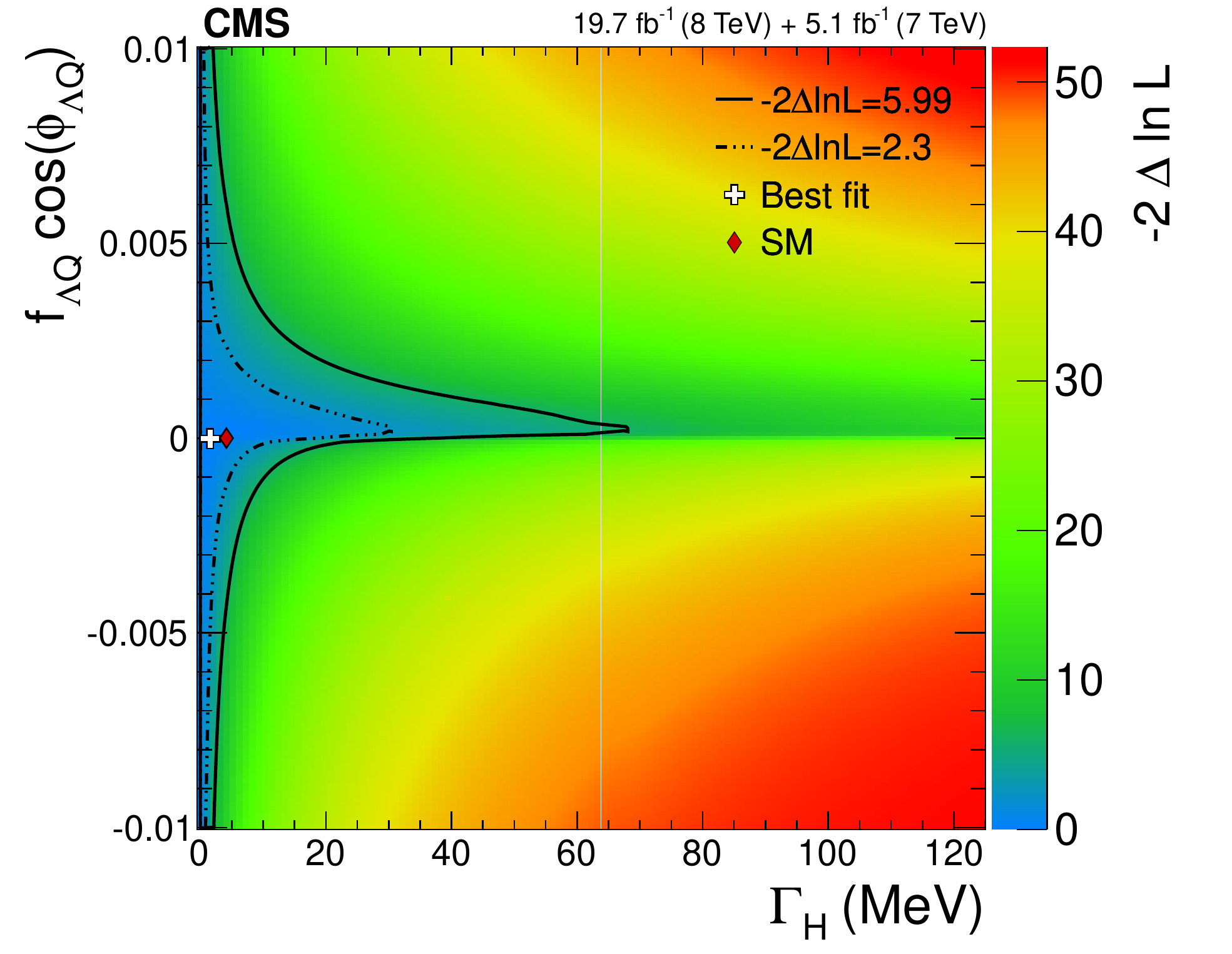}
\includegraphics[width=\cmsFigWidth]{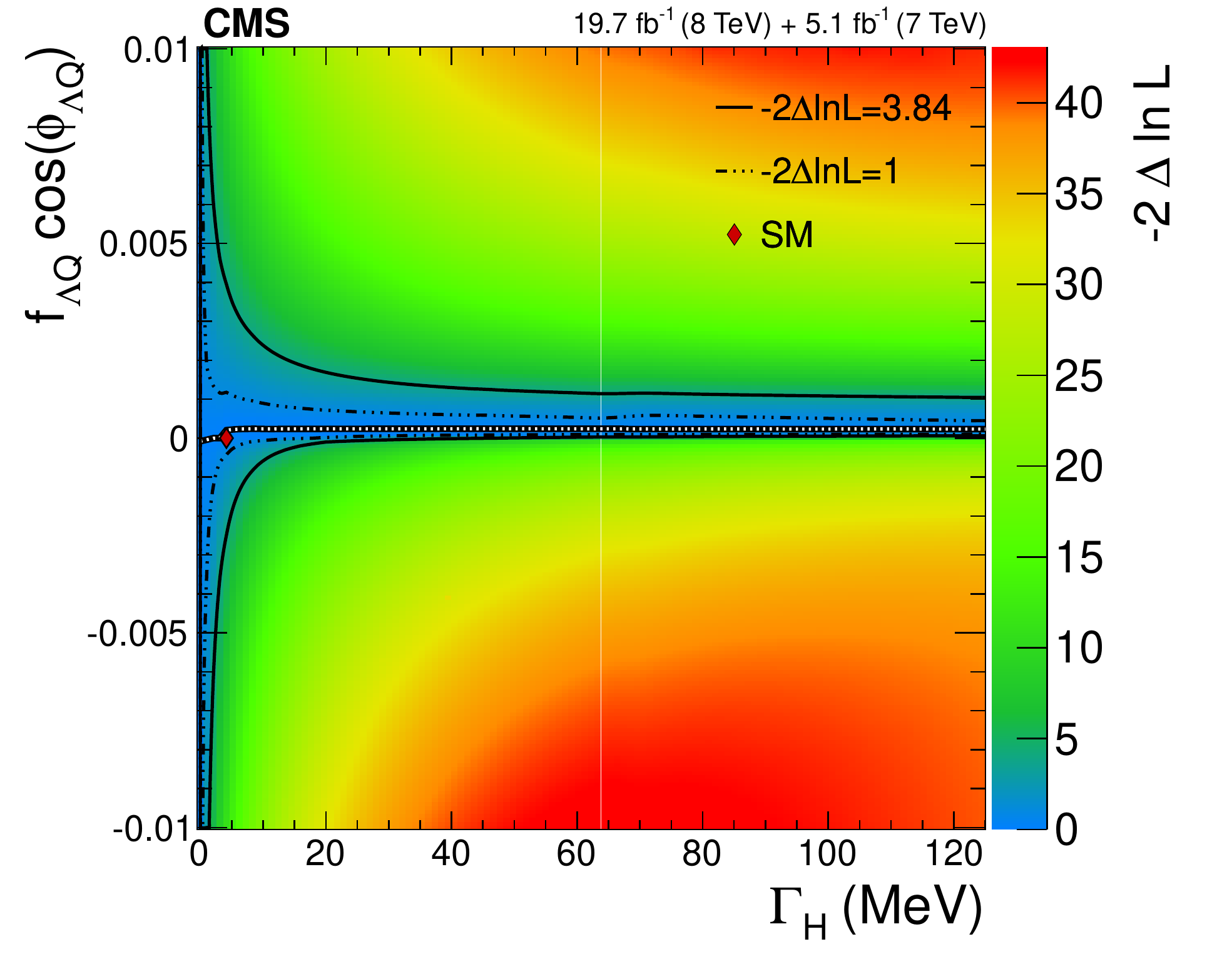}
\caption{
Observed distribution of $-2\ln(\mathcal{L}/\mathcal{L}_\text{max})$ as a function of $\GH$ and $\fLQ\cos \phiLQ$
with the assumption $\phiLQ=0$ or $\pi$ (top panel).
The bottom panel shows the observed conditional likelihood scan as a function of $\fLQ\cos \phiLQ$ for a given $\GH$.
The likelihood contours are shown for the two-parameter 68\% and 95\% \CLs (top)
and for the one-parameter 68\% and 95\% \CLs (bottom). The black curve with white dots on the bottom panel
shows the $\fLQ\cos \phiLQ$ minima at each $\GH$ value.
}
\label{fig:2Dlimit}
\end{figure}

\begin{figure}[h!tb]
\centering
\includegraphics[width=\cmsFigWidth]{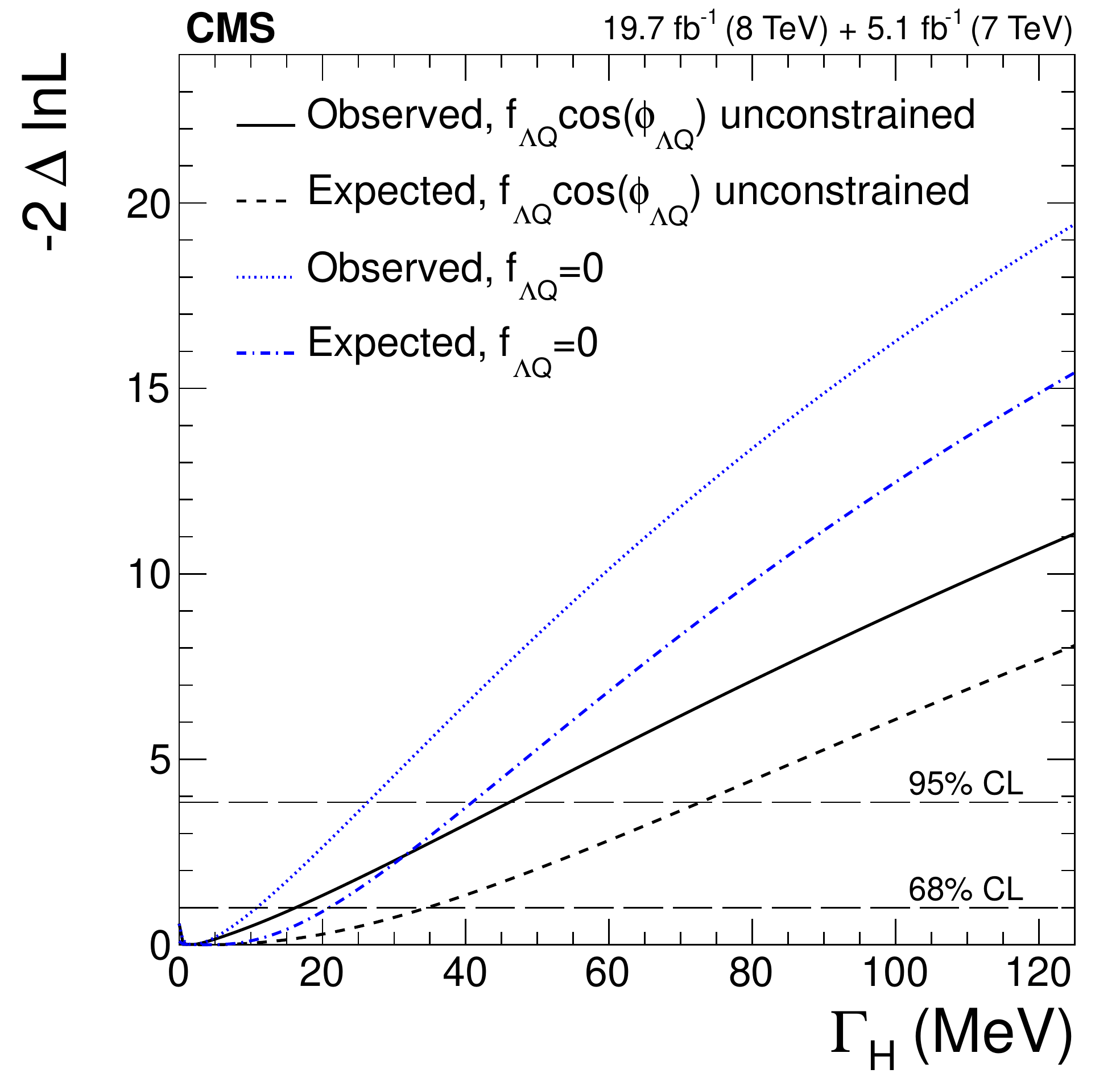}
\includegraphics[width=\cmsFigWidth]{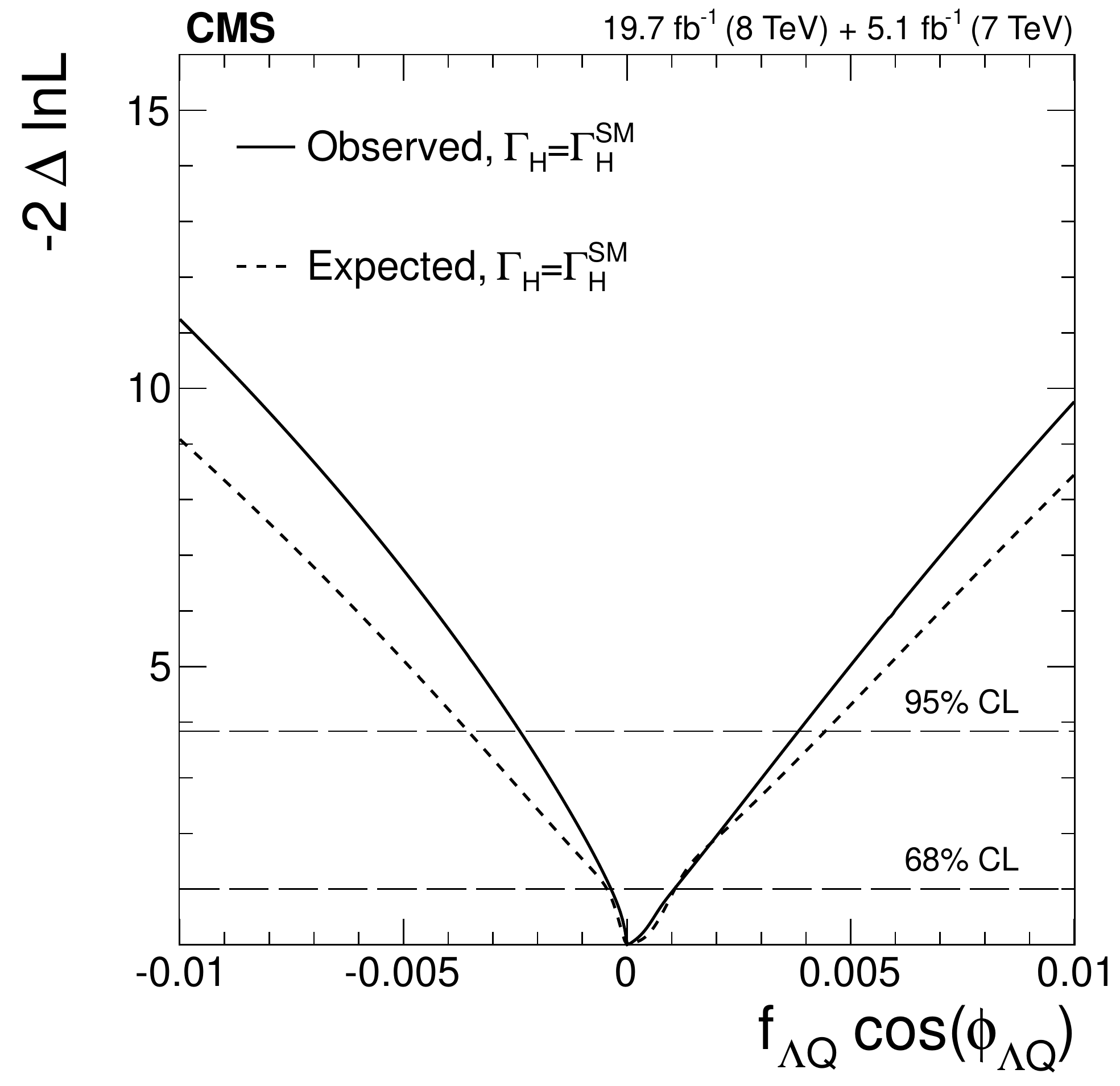}
\caption{
Observed (solid) and expected (dashed) distributions of $-2\ln(\mathcal{L}/\mathcal{L}_\text{max})$ as a function of
$\GH$  (top) and $\fLQ\cos\phiLQ$ (bottom).
On the top panel, the $\fLQ$ value is either constrained to zero (blue) or left unconstrained (black, weaker limit),
while $\GH=\GHSM$ and $\phiLQ=0$ or $\pi$ are assumed on the bottom.
}
\label{fig:1Dlimit}
\end{figure}

The allowed 68\% and 95\% \CL intervals are defined using the profile likelihood function values $-2\ln(\mathcal{L}/\mathcal{L}_\text{max}) = 2.30$ and 5.99, respectively, for the two-parameter constraints presented, and $-2\ln(\mathcal{L}/\mathcal{L}_\text{max}) = 1.00$ and 3.84, respectively, for the one-parameter constraints. Exact coverage is expected in the asymptotic limit~\cite{wilks1938}, and the approximate coverage has been tested at several different parameter values with the quoted results having been found to be conservative.
The observed distribution of the likelihood as a two-parameter function of \GH\ and $\fLQ \cos \phiLQ$, with $\phiLQ=0$ or $\pi$,
is shown in Fig.~\ref{fig:2Dlimit}. Also shown is the one-parameter, conditional likelihood scan of $\fLQ\cos \phiLQ$ for a given $\GH$,
where the $-2\ln(\mathcal{L}/\mathcal{L}_\text{max})$ distribution is shown for $\mathcal{L}_\text{max}$
adjusted according to the most likely value of $\fLQ \cos \phiLQ$ at the given value of $\GH$.
The observed and expected likelihood distributions as a function of \GH\ are shown in Fig.~\ref{fig:1Dlimit},
where $\fLQ$ is either constrained to zero or left unconstrained.
The observed (expected) central values with 68\% \CL uncertainties are
$\GH=2^{+9}_{-2}~(4^{+17}_{-4})$\MeV with $\fLQ=0$,  and
$\GH=2^{+15}_{-2}~(4^{+30}_{-4})$\MeV with $\fLQ$ unconstrained and $\phiLQ=0$ or $\pi$.
The observed (expected) constraints at 95\% \CL are
$\GH<26\,(41)$\MeV with $\fLQ=0$, and
$\GH<46\,(73)$\MeV with $\fLQ$ unconstrained and $\phiLQ=0$ or $\pi$.
These observed (expected) upper limits on the \PH boson width at 95\% \CL correspond
through Eq.~(\ref{eq:tauHgammaH}) to
the lower limits on the \PH boson average lifetime
$\tau_\PH>2.5\times10^{-8}\,(1.6\times10^{-8})$\unit{fs} with $\fLQ=0$ and
$\tau_\PH>1.4\times10^{-8}\,(9\times10^{-9})$\unit{}fs with $\fLQ$ unconstrained and $\phiLQ=0$ or $\pi$.

The result with the constraint $\fLQ=0$ is consistent with the earlier one from the $\PH\to\Z\Z\to4\ell$ channel~\cite{CMS-HIG-14-002}. It can be reinterpreted as an off-shell signal strength with the change of parameters
$\mu^\text{off-shell}_{\text{vv}\PH} = \mu_{\text{vv}\PH} \, \GH/\GHSM$, provided the signal strength $\mu_{\text{vv}\PH}$ for the on-shell region is uncorrelated with the signal strength $\mu^\text{off-shell}_{\text{vv}\PH}$ for the off-shell region in the likelihood scan.
The observed (expected) central values and the 68\% \CL uncertainties of $\GH$ with the $\fLQ=0$ constraint correspond to
$\mu^\text{off-shell}_{\Pg\Pg\PH}=0.5^{+2.2}_{-0.5}~(1.0^{+5.1}_{-1.0})$ and
$\mu^\text{off-shell}_{\V\V\PH}=0.4^{+10.5}_{-0.4}~(1.0^{+20.6}_{-1.0})$,
and the observed (expected) constraints at 95\% \CL become
$\mu^\text{off-shell}_{\Pg\Pg\PH}<6.2\,(9.3)$ and
$\mu^\text{off-shell}_{\V\V\PH}<31.3\,(44.4)$.
There is no constraint on the ratio $\mu^\text{off-shell}_{\V\V\PH}/\mu^\text{off-shell}_{\Pg\Pg\PH}$ at 68\% \CL.
The $\GH$ limits with $\fLQ$ unconstrained are weaker because a small nonzero value $\fLQ \sim 2 \times 10^{-4}$ leads to
destructive interference between the $a_1$ and $\Lambda_Q$ terms in Eq.~(\ref{eq:fullampl-formfact-spin0}) when $\phiLQ=0$.
This interference reduces the expected signal yield at these parameter values, thereby reducing the exclusion power for $\GH>\GHSM$. This effect is also illustrated in Fig.~\ref{fig:massfull}.

No constraint on $\fLQ$ can be obtained in the limit $\GH\to 0$ because, as displayed in Fig.~\ref{fig:2Dlimit}, the number of expected
off-shell events vanishes.
The constraints on $\fLQ \cos \phiLQ$ given particular $\GH$ values become tighter for increasing $\GH$.
The limits on $\fLQ \cos \phiLQ$ with the assumption $\GH=\GHSM$ are presented in Fig.~\ref{fig:1Dlimit}.
The observed (expected) value is
$\fLQ \cos \phiLQ=0^{+1.0}_{-0.4}\, (0^{+1.1}_{-0.4})\, \times 10^{-3}$,
and the allowed region at 95\% \CL is $[-2.4,~3.8]\,\times\,10^{-3}~([-3.6,~4.4]\,\times\,10^{-3})$.

\section{Conclusions} \label{sec:Conclusions}

Constraints on the lifetime and the width of the \PH boson are obtained from
$\PH\to\Z\Z\to 4\ell$ events using the data recorded by the CMS experiment during the LHC run~1.
The measurement of the \PH boson lifetime is derived from its flight distance
in the CMS detector with the upper bound ${\tau_\PH<190}$\unit{fs} at the 95\% \CL,
corresponding to a lower bound on the width $\GH > 3.5 \times 10^{-9}$\MeV.
The measurement of the width is obtained from an off-shell production technique,
generalized to include additional anomalous couplings of the \PH boson to two electroweak bosons.
This measurement provides a joint constraint on the \PH boson width and a parameter that
quantifies an anomalous coupling contribution through an on-shell cross-section fraction $\fLQ$.
The observed limit on the \PH boson width is ${\GH<46}$\MeV at the 95\% \CL with $\fLQ$ left unconstrained
while it is $\GH<26$\MeV at the 95\% \CL for $\fLQ=0$.
The constraint $\fLQ < 3.8\times10^{-3}$ at the 95\% \CL is obtained assuming the \PH boson width expected
in the SM, and the $\fLQ$ constraints given any other width value are also presented.
Table~\ref{tab:width_lifetime_summary} summarizes the width and corresponding lifetime limits,
and Table~\ref{tab:fLQ_summary} summarizes the limits on $\fLQ$ under the different $\phiLQ$ scenarios
that can be interpreted from this analysis, and provides the corresponding limits on ${\sqrt{a_1} \, \Lambda_Q}$.

\begin{table*}[htb]
\centering
\topcaption{
Observed and expected allowed intervals at the 95\% \CL on the $\PH$ boson average lifetime $\tau_\PH$ and width $\GH$ obtained combining the width and lifetime analyses. The constraints are separated into the two conditions
used in the width measurement, with either the constraint $\fLQ=0$, or $\fLQ$ left unconstrained and $\phiLQ=0$ or $\pi$. The upper (lower) limits on $\PH$ boson average lifetime $\tau_\PH$ are related to the lower (upper) limits on
$\PH$ boson width $\GH$ through Eq.~(\ref{eq:tauHgammaH}).
\label{tab:width_lifetime_summary}
}
\begin{scotch}{ccc{c}@{\hspace*{5pt}}cc}
\multirow{2}{*}{Parameter} &  \multicolumn{2}{c}{$\fLQ=0$}
 && \multicolumn{2}{c}{$\fLQ$ unconstrained, $\phiLQ=0$ or $\pi$} \\ \cline{2-3}\cline{5-6}
 & Observed & Expected
 && Observed & Expected \\
\hline\\ [-2.2ex]
$\tau_\PH$ (fs)
& $[2.5\times10^{-8},~190]$ & $[1.6\times10^{-8},~190]$
&& $[1.4\times10^{-8},~190]$ & $[9\times10^{-9},~190]$ \\
$\GH$ (\MeVns{})
& $[3.5\times10^{-9},~26]$ & $[3.6\times10^{-9},~41]$
&& $[3.5\times10^{-9},~46]$ & $[3.6\times10^{-9},~73]$ \\
\end{scotch}
\end{table*}
\begin{table*}[htb]
\centering
\topcaption{
Observed and expected allowed intervals at the 95\% \CL on the $\fLQ$ on-shell effective cross-section fraction and its interpretation in terms of the anomalous
coupling parameter $\Lambda_Q$ assuming $\GH=\GHSM$. Results are presented assuming either $\phiLQ=0$ or $\phiLQ=\pi$. The allowed intervals on $\fLQ$ are also translated to the equivalent quantity $\sqrt{a_1} \,\Lambda_Q$ through Eq.~(\ref{eq:fLQ_definition}), where the coefficient $a_1$ is allowed to be different from its SM value $a_1=2$.
\label{tab:fLQ_summary}
}
\begin{scotch}{ccc{c}@{\hspace*{5pt}}cc}
\multirow{2}{*}{Parameter}                &  \multicolumn{2}{c}{$\phiLQ=0$}
 && \multicolumn{2}{c}{$\phiLQ=\pi$} \\ \cline{2-3} \cline{5-6}
 & Observed & Expected
 && Observed & Expected \\
\hline\\ [-2.2ex]
$\fLQ$ & ${<}3.8\times10^{-3}$ & ${<}4.4\times10^{-3}$ && ${<}2.4\times10^{-3}$ & ${<}3.6\times10^{-3}$ \\
$\sqrt{a_1} \, \Lambda_Q\, (\GeVns{})$  & $>$500 & $>$490  && $>$570 & $>$510 \\
\end{scotch}
\end{table*}

\ifthenelse{\boolean{cms@external}}{}{\clearpage}

\begin{acknowledgments}\label{sec:Acknowledgments}
We thank Markus Schulze for optimizing the \textsc{JHUGen} Monte Carlo simulation program and matrix element library for this analysis. We congratulate our colleagues in the CERN accelerator departments for the excellent performance of the LHC and thank the technical and administrative staffs at CERN and at other CMS institutes for their contributions to the success of the CMS effort. In addition, we gratefully acknowledge the computing centers and personnel of the Worldwide LHC Computing Grid for delivering so effectively the computing infrastructure essential to our analyses. Finally, we acknowledge the enduring support for the construction and operation of the LHC and the CMS detector provided by the following funding agencies: the Austrian Federal Ministry of Science, Research and Economy and the Austrian Science Fund; the Belgian Fonds de la Recherche Scientifique, and Fonds voor Wetenschappelijk Onderzoek; the Brazilian funding agencies (CNPq, CAPES, FAPERJ, and FAPESP); the Bulgarian Ministry of Education and Science; CERN; the Chinese Academy of Sciences, Ministry of Science and Technology, and National Natural Science Foundation of China; the Colombian Funding Agency (COLCIENCIAS); the Croatian Ministry of Science, Education and Sport, and the Croatian Science Foundation; the Research Promotion Foundation, Cyprus; the Ministry of Education and Research, Estonian Research Council via IUT23-4 and IUT23-6 and European Regional Development Fund, Estonia; the Academy of Finland, Finnish Ministry of Education and Culture, and Helsinki Institute of Physics; the Institut National de Physique Nucl\'eaire et de Physique des Particules~/~CNRS, and Commissariat \`a l'\'Energie Atomique et aux \'Energies Alternatives~/~CEA, France; the Bundesministerium f\"ur Bildung und Forschung, Deutsche Forschungsgemeinschaft, and Helmholtz-Gemeinschaft Deutscher Forschungszentren, Germany; the General Secretariat for Research and Technology, Greece; the National Scientific Research Foundation, and National Innovation Office, Hungary; the Department of Atomic Energy and the Department of Science and Technology, India; the Institute for Studies in Theoretical Physics and Mathematics, Iran; the Science Foundation, Ireland; the Istituto Nazionale di Fisica Nucleare, Italy; the Ministry of Science, ICT and Future Planning, and National Research Foundation (NRF), Republic of Korea; the Lithuanian Academy of Sciences; the Ministry of Education, and University of Malaya (Malaysia); the Mexican Funding Agencies (CINVESTAV, CONACYT, SEP, and UASLP-FAI); the Ministry of Business, Innovation and Employment, New Zealand; the Pakistan Atomic Energy Commission; the Ministry of Science and Higher Education and the National Science Centre, Poland; the Funda\c{c}\~ao para a Ci\^encia e a Tecnologia, Portugal; JINR, Dubna; the Ministry of Education and Science of the Russian Federation, the Federal Agency of Atomic Energy of the Russian Federation, Russian Academy of Sciences, and the Russian Foundation for Basic Research; the Ministry of Education, Science and Technological Development of Serbia; the Secretar\'{\i}a de Estado de Investigaci\'on, Desarrollo e Innovaci\'on and Programa Consolider-Ingenio 2010, Spain; the Swiss Funding Agencies (ETH Board, ETH Zurich, PSI, SNF, UniZH, Canton Zurich, and SER); the Ministry of Science and Technology, Taipei; the Thailand Center of Excellence in Physics, the Institute for the Promotion of Teaching Science and Technology of Thailand, Special Task Force for Activating Research and the National Science and Technology Development Agency of Thailand; the Scientific and Technical Research Council of Turkey, and Turkish Atomic Energy Authority; the National Academy of Sciences of Ukraine, and State Fund for Fundamental Researches, Ukraine; the Science and Technology Facilities Council, UK; the U.S. Department of Energy, and the U.S. National Science Foundation.

Individuals have received support from the Marie Curie program and the European Research Council and EPLANET (European Union); the Leventis Foundation; the A. P. Sloan Foundation; the Alexander von Humboldt Foundation; the Belgian Federal Science Policy Office; the Fonds pour la Formation \`a la Recherche dans l'Industrie et dans l'Agriculture (FRIA-Belgium); the Agentschap voor Innovatie door Wetenschap en Technologie (IWT-Belgium); the Ministry of Education, Youth and Sports (MEYS) of the Czech Republic; the Council of Science and Industrial Research, India; the HOMING PLUS program of the Foundation for Polish Science, cofinanced from European Union, Regional Development Fund; the Compagnia di San Paolo (Torino); the Consorzio per la Fisica (Trieste); MIUR project 20108T4XTM (Italy); the Thalis and Aristeia programs cofinanced by EU-ESF and the Greek NSRF; the National Priorities Research Program by Qatar National Research Fund; the Rachadapisek Sompot Fund for Postdoctoral Fellowship, Chulalongkorn University (Thailand); and the Welch Foundation.
\end{acknowledgments}

\bibliography{auto_generated}

\cleardoublepage \appendix\section{The CMS Collaboration \label{app:collab}}\begin{sloppypar}\hyphenpenalty=5000\widowpenalty=500\clubpenalty=5000\input{HIG-14-036-authorlist.tex}\end{sloppypar}
\end{document}

%% file: HIG-14-036-authorlist.tex
\textbf{Yerevan Physics Institute,  Yerevan,  Armenia}\\*[0pt]
V.~Khachatryan, A.M.~Sirunyan, A.~Tumasyan
\vskip\cmsinstskip
\textbf{Institut f\"{u}r Hochenergiephysik der OeAW,  Wien,  Austria}\\*[0pt]
W.~Adam, E.~Asilar, T.~Bergauer, J.~Brandstetter, E.~Brondolin, M.~Dragicevic, J.~Er\"{o}, M.~Flechl, M.~Friedl, R.~Fr\"{u}hwirth\cmsAuthorMark{1}, V.M.~Ghete, C.~Hartl, N.~H\"{o}rmann, J.~Hrubec, M.~Jeitler\cmsAuthorMark{1}, V.~Kn\"{u}nz, A.~K\"{o}nig, M.~Krammer\cmsAuthorMark{1}, I.~Kr\"{a}tschmer, D.~Liko, T.~Matsushita, I.~Mikulec, D.~Rabady\cmsAuthorMark{2}, B.~Rahbaran, H.~Rohringer, J.~Schieck\cmsAuthorMark{1}, R.~Sch\"{o}fbeck, J.~Strauss, W.~Treberer-Treberspurg, W.~Waltenberger, C.-E.~Wulz\cmsAuthorMark{1}
% ...
\vskip\cmsinstskip
% ...
1:~~Also at Vienna University of Technology, Vienna, Austria\\
2:~~Also at CERN, European Organization for Nuclear Research, Geneva, Switzerland\\
% ...
